\documentclass[usenatbib]{mn2e}
\usepackage{natbib}
\usepackage{amsmath}
\usepackage{epsfig}

\def\reff@jnl#1{{\rm#1\/}}

\def\aj{\reff@jnl{AJ}}                  
\def\araa{\reff@jnl{ARA\&A}}            
\def\apj{\reff@jnl{ApJ}}                        
\def\apjl{\reff@jnl{ApJ}}               
\def\apjs{\reff@jnl{ApJS}}              
\def\apss{\reff@jnl{Ap\&SS}}            
\def\aap{\reff@jnl{A\&A}}               
\def\aapr{\reff@jnl{A\&A~Rev.}}         
\def\aaps{\reff@jnl{A\&AS}}             
\def\baas{\reff@jnl{BAAS}}              
\def\jrasc{\reff@jnl{JRASC}}            
\def\memras{\reff@jnl{MmRAS}}           
\def\mnras{\reff@jnl{MNRAS}}            
\def\physrep{\reff@jnl{Phys.Rep.}}
\def\pra{\reff@jnl{Phys.Rev.A}}         
\def\prb{\reff@jnl{Phys.Rev.B}}         
\def\prc{\reff@jnl{Phys.Rev.C}}         
\def\prd{\reff@jnl{Phys.Rev.D}}         
\def\prl{\reff@jnl{Phys.Rev.Lett}}      
\def\pasp{\reff@jnl{PASP}}              
\def\pasj{\reff@jnl{PASJ}}              
\def\skytel{\reff@jnl{S\&T}}            
\def\solphys{\reff@jnl{Solar~Phys.}}    
\def\sovast{\reff@jnl{Soviet~Ast.}}     
\def\ssr{\reff@jnl{Space~Sci.Rev.}}     
\def\nat{\reff@jnl{Nature}}             

\def\sun{\hbox{$\odot$}}

\newcommand{\hmpc}{\ensuremath{h^{-1}\mathrm{Mpc}}}
\newcommand{\hkpc}{\ensuremath{h^{-1}\mathrm{kpc}}}
\newcommand{\hMsun}{h^{-1}M_{\odot}}

\newcommand{\ds}{\ensuremath{\Delta\Sigma}}
\newcommand{\scinv}{\ensuremath{\Sigma_c^{-1}}}

\newcommand{\hinvk}{$h^{-1}$kpc}
\newcommand{\avgnm}{\ensuremath{\langle N(M)\rangle}}

\def\beq{\begin{equation}}
\def\eeq{\end{equation}}

\title[Star formation]{Galaxy halo masses and 
satellite fractions 
from galaxy-galaxy lensing in the SDSS: 
stellar mass, luminosity, morphology and environment 
dependencies 
}  

\author[Mandelbaum et. al.]{
Rachel Mandelbaum$^1$\thanks{Electronic address:
    {\tt rmandelb@princeton.edu}},
Uro\v{s} Seljak$^{1,2}$, Guinevere Kauffmann$^3$,\newauthor
Christopher M. Hirata$^4$, Jonathan Brinkmann$^5$ 
\\$^1$Department of Physics, Jadwin Hall, Princeton University,
      Princeton NJ 08544, USA
\\$^2$International Centre for Theoretical Physics, Strada Costiera 11,
      34014 Trieste, Italy
\\$^3$Max-Planck-Institut f\"{u}r Astrophysik,
      Karl-Schwarzschild-Strasse 1, 85748 Garching, Germany
\\$^4$Institute for Advanced Study, Einstein Drive,
      Princeton, NJ 08540, USA
\\$^5$Apache Point Observatory, 2001 Apache Point Road,
      Sunspot NM 88349, USA
}
\date{\today}

\begin{document}
\maketitle

\begin{abstract}
The relationship between galaxies and dark matter can be characterized
by the halo mass of the central galaxy and the fraction of 
galaxies that are satellites.  Here we present observational constraints 
from the SDSS on these quantities as a function of $r$-band luminosity
and stellar mass using galaxy-galaxy weak lensing, with a total of 351~507
lenses. We use stellar masses derived from spectroscopy  
and virial halo masses derived from weak gravitational lensing to
determine the efficiency with which baryons in the halo of the central
galaxy have been converted into stars. 
We find that an
$L_*$ galaxy with a stellar mass of $6\times 10^{10}M_{\sun}$ is
hosted by a halo with mass of $1.4\times 10^{12}h^{-1}M_{\sun}$, independent
of morphology, yielding baryon conversion efficiencies of $17_{-5}^{+10}$
(early types) and $16_{-6}^{+15}$ (late types) per cent at the 95 per
cent CL (statistical, not including systematic uncertainty due to
assumption of a universal initial mass function, or IMF). 
We find that for a given stellar mass, the halo mass is
independent of morphology below $M_{stellar}=10^{11}M_{\sun}$, 
in contrast to typically a factor of two difference in halo mass 
between ellipticals and spirals at a fixed luminosity. This suggests that
stellar mass is a good  proxy for halo mass in this range and 
should be used preferentially whenever a halo mass selected 
sample is needed. For higher stellar masses, the conversion efficiency is a
declining function of stellar mass, and the differences in halo mass
between early and late types become larger, reflecting the fact that 
most group and cluster halos with masses above $10^{13}M_{\sun}$
host ellipticals at the center, while even the brightest 
central spirals are hosted by halos of mass below $10^{13}M_{\sun}$. 
We find that the fraction of spirals that are satellites is roughly 10-15 per
cent independent of stellar mass or luminosity,  while for 
ellipticals this fraction decreases with stellar mass from 50 per cent
at $10^{10}M_{\sun}$ to 10 per cent at $3 \times 10^{11}M_{\sun}$ or
20 per cent at the maximum luminosity considered.
We split the elliptical sample by local density, and find that at a given
luminosity there is no difference in the signal on scales below 100 \hkpc{}
between high and low density regions, suggesting that tidal stripping
inside large halos does not remove most of the dark matter from the
early type satellites.    This result is dominated by halos in the
mass range $10^{13}-10^{14} \hMsun$, and is an average over all
separations from the group or cluster center.
\end{abstract}


\section{Introduction}\label{S:intro}

The connection between the spatial distribution of galaxies
and dark matter (DM) is an essential ingredient
in the physics of galaxy formation. One very useful probe of the
galaxy-DM connection that recently became available is weak lensing around
galaxies, or galaxy-galaxy (hereinafter g-g) lensing
 \citep{1984ApJ...281L..59T,1996ApJ...466..623B,1998ApJ...503..531H,2000AJ....120.1198F,2001astro.ph..8013M,2001ApJ...551..643S,2003MNRAS.340..609H,2004ApJ...606...67H,2004AJ....127.2544S,2005MNRAS.361.1287M,2005PhRvD..71d3511S}.  Gravitational lensing induces tangential shear 
distortions of background galaxies around foreground galaxies, allowing
direct measurement of the galaxy-DM correlation function around
galaxies.  The individual distortions are small (of order 0.1\%), but by
averaging over all foreground galaxies within a given subsample, we
obtain high signal to noise in the shear as a function of angular
separation from the galaxy.  If we know the lens redshifts, the shear signal
can be related to the projected mass density as a function of proper
distance from the galaxy.  This allows us to determine 
the averaged DM distribution around any given galaxy sample.

In recent years, the progress on the observational side of g-g lensing
has been remarkable.  In the latest Sloan Digital Sky Survey (SDSS)
analyses \citep{2004AJ....127.2544S,2005PhRvD..71d3511S},
20--$30\sigma$ detections of the signal as a function of physical
separation have been obtained.  Similarly high S/N detections 
have also been observed as a function of angular separation
with other surveys \citep{2004ApJ...606...67H}, but the ability to use
spectroscopic 
redshifts for lenses is a major advantage to doing lensing with the
SDSS. The high statistical
power has been accompanied by a more careful investigation of systematic
errors, such as calibration biases and intrinsic alignments,
which for the SDSS are currently around 10 per cent and therefore
already dominate the error budget \citep{2005MNRAS.361.1287M}.

In this work, we seek to use g-g weak lensing to explore the
galaxy-DM connection for particular subsamples of lenses.  By
comparison with the predicted signal from a halo model, as done for
simulations in \cite{2005MNRAS.362.1451M}, we can extract
average central halo masses and satellite fractions.  These
calculations are done as a function of morphology and of environment,
in samples selected based on both stellar masses and luminosity.  We
expect that the divisions by morphology and by environment may be
related, due to the relationships between color and environment, with
red galaxies typically found in overdense regions
\citep{1976ApJ...208...13D,1980ApJ...236..351D,1984ApJ...281...95P,1998ApJ...504L..75B,1999ApJ...527...54B,2001ApJ...563..736C,2003AAS...20314501B,2003ApJ...585L...5H,2004ApJ...615L.101B,2004MNRAS.348.1355B,2004ApJ...601L..29H,2005MNRAS.356.1155C}.
 By determining average central halo masses and satellite fractions as a
 function of these parameters, we hope to gain some insight into
 processes of galaxy formation, and ultimately into the galaxy-DM
 connection.  We note that this approach has been used before, 
 by \citet{2002MNRAS.335..311G} based on data in \cite{2001astro.ph..8013M}, 
but with a simpler form of the
 halo model, with a much smaller sample of lenses so lower statistical
 power, and only using the luminosities, not
 stellar masses.  Due to our larger sample of lenses, our
 better-understood calibration, and our
 inclusions of stellar masses which are better tracers of stellar
and dark matter
 content than luminosities, this work constitutes a significant
 improvement over that one.  Another recent work,
 \citet{2005ApJ...635...73H}, used 
 stellar masses for RCS data derived from $B-V$ colors from CFHT
 photometry in order to derive halo masses as a function of
 luminosity, and star formation efficiencies as a
 function of morphology, but used only isolated lenses and thus
 did not derive satellite fractions.  Furthermore, the lack of
 spectroscopic redshifts for lenses in that work, which allow the derivation of
 stellar masses via spectral indicators and the computation of the
 lensing signal as a function of transverse separation in this work,
 complicates the  analysis.  Halo model analysis of galaxy-galaxy 
 autocorrelations has been 
 done observationally by several groups
 \citep{2003MNRAS.340..771V,2005astro.ph..9033C,2005ApJ...630....1Z}, and this 
 halo model analysis of galaxy-DM cross-correlations is, in
 many ways, complementary to that approach.

We begin by introducing the g-g lensing formalism and the halo model
that is used to extract 
information about central halo masses and satellite fractions  in
\S\ref{S:hm}.  \S\ref{S:data} includes a description of the SDSS data
used for this analysis.  We present the
lensing signal and the halo model fits in \S\ref{S:results}, and
interpretation of these 
results.  We conclude in \S\ref{S:conclusions} with a summary of our findings.

Here we note the cosmological model and units used in this paper.
All computations assume a flat $\Lambda$CDM universe with
$\Omega_m=0.3$, $\Omega_{\Lambda}=0.7$, and $\sigma_8=0.9$.  Distances
quoted for 
transverse lens-source separation are comoving (rather than physical)
\hinvk, where $H_0=100\,h$ km$\mathrm{s}^{-1}\,\mathrm{Mpc}^{-1}$.
Likewise, \ds{} is computed using the expression for \scinv{} in
comoving coordinates, Eq.~\ref{E:sigmacrit}.  In the units
used, $H_0$ scales out of everything, so our results are independent of
this quantity.  All confidence intervals in the text and tables are 95
per cent confidence level ($2\sigma$) unless explicitly noted otherwise.

\section{Weak lensing formalism and halo model}\label{S:hm}

Galaxy-galaxy weak lensing provides a simple way to probe the
connection between galaxies and matter via their
cross-correlation function
\beq
\xi_{g,m}(\vec{r}) = \langle \delta_g (\vec{x})
\delta_{m}^{*}(\vec{x}+\vec{r})\rangle 
\eeq
where $\delta_g$ and $\delta_{m}$ are overdensities of galaxies and
matter, respectively.  This cross-correlation can be related to the
projected surface density
\beq\label{E:sigmar}
\Sigma(R) = \overline{\rho} \int \left[1+\xi_{g,m}\left(\sqrt{R^2 + \chi^2}\right)\right] d\chi
\eeq
(where $r^2=R^2+\chi^2$), which is then related to the observable
quantity for lensing, 
\beq\label{E:ds}
\ds(R) = \gamma_t(R) \Sigma_c= \overline{\Sigma}(<R) - \Sigma(R), 
\eeq
where the second relation is true only for a matter distribution that
is axisymmetric along the line of sight.  This observable quantity can
be expressed as the product of two factors, a tangential shear
$\gamma_t$ and a geometric factor
\beq\label{E:sigmacrit}
\Sigma_c = \frac{c^2}{4\pi G} \frac{D_S}{D_L D_{LS}(1+z_L)^2}
\eeq
where $D_L$ and $D_S$ are angular diameter distances to the lens and
source, $D_{LS}$ is the angular diameter distance between the lens
and source, and the factor of $(1+z_L)^{-2}$ arises due to our use of
comoving coordinates.  For a given lens redshift,
$\Sigma_c^{-1}$ rises from zero at $z_s = z_L$ to an asymptotic value
at $z_s \gg z_L$; that asymptotic value is an increasing function of
lens redshift.

There are two basic approaches that can be used to extract information about
properties of the galaxy distribution (e.g., halo masses and satellite
fractions) from the g-g lensing signal.  The first approach
is to compare directly against N-body simulations
\citep{2001MNRAS.321..439G,2003MNRAS.339..387Y,2004ApJ...614..533T,2004ApJ...601....1W}.
While this approach has the advantage of being 
fairly direct, it has the disadvantage that even assuming a given
cosmological model, the process of galaxy formation is not
sufficiently understood to result in unique predictions for that model.  When
combined with the fact that the cosmological model is not itself
fully determined yet, one would have to expend tremendous computational
resources to run multiple simulations with different cosmologies and models.
Furthermore, current simulations still suffer from a limited 
dynamical range, in the sense that they require a high mass and force 
resolution to resolve individual galaxies and their associated DM 
halos, while at the same time they must also have sufficiently large
volume to simulate a representative region of the universe. Several
simulations of varying box size are thus needed to cover the whole
observational range in luminosity and scale.  

Another approach to model the relation between galaxies and 
dark matter is to use a halo model
\citep[e.g.][]{2000MNRAS.318.1144P,2000MNRAS.318..203S,2001ApJ...546...20S,2002PhR...372....1C}, which in this application can be used as a 
phenomenological description of the processes that determine
the lensing signal for particular types of galaxies.  Comparison of the
halo model predicted signal versus the real signal can lead to the
determination of quantities such as the virial mass distribution and
the fraction of these galaxies that are 
satellites, which are useful quantities for constraining 
the galaxy formation models and cosmological models.  This approach
has the advantage that it is not as computationally expensive, so
large areas of parameter space can be explored very quickly, but must be
compared against simulations to ensure that it works properly. 
\cite{2005MNRAS.362.1451M} compared the halo model 
with simulations, and determined that central halo masses ($M_{cent}$)
and satellite fractions ($\alpha$) can be extracted adequately from the lensing
signal, provided that the distribution of central halo masses is not
too broad ($\sigma_{M}/M_{cent}$ less than about a factor of
two), or that corrections be applied if it is 
broad.  Thus, in this paper we use those results to extract information
from the measured weak lensing signal from the SDSS using halo model
fits, as has already been done by \cite{2005PhRvD..71d3511S}
(including variation of the results with $\Omega_m$ and $\sigma_8$,
for which we refer 
the reader to that paper).
The halo model used here is the same as that from
  \cite{2005MNRAS.362.1451M}, to which the interested reader can refer
  for details.  Here we give only the basic details necessary to
  understand this paper.

The halo model can be used to derive the galaxy-DM cross-power spectrum
$P_{g,dm}(k)$ by considering separately the contributions of central
galaxies (which lie in halos that are not contained within another
halo) and satellite galaxies (which lie in subhalos contained entirely
within another halo).  There are then one-halo or Poisson terms,
derived from the cross-correlation of the galaxy with its own matter
distribution (for central galaxies and satellites)  and
with that of the host halo (for satellites), and a halo-halo (h-h) term derived
from the cross-correlation with the mass of other halos.  The latter
is negligible on the small ($<2\,h^{-1}$Mpc) scales to which we limit
ourselves in this work. Therefore, rather
than attempting to fit for it by fitting to the bias, we fix the
bias by doing the halo model fits without the h-h term to get
the halo mass, then using $b(M)$ from \cite{2004MNRAS.355..129S} to
redo the fits for $M_{cent}$ and $\alpha$ with
the appropriate value of bias fixing the h-h term.  
Including this correction makes very little difference on the 
final results, as expected since on small scales the h-h term is 
negligible. For the same reason we ignore the morphology 
dependence of halo bias, suggested by 
recent simulations \cite{2005MNRAS.363L..66G} that studied the
dependence of clustering on halo formation time. 
For splits by local density, we
use the appropriate $b(M)$ for that mass for the high-density samples
(again, ignoring dependence of bias on local environment), and
do not include a h-h term at all for low-density samples.  

The one-halo term requires various ingredients, such as the halo mass
function $dn/dM$, 
the radial profile of dark matter within halos $\rho(r)$, the radial distribution
of galaxies within groups and clusters, the conditional halo mass
probability distribution, and the tidal stripping of satellites in clusters.
The DM profiles are assumed here to be 
NFW profiles \citep{1996ApJ...462..563N} with concentration parameter
\citep{2001MNRAS.321..559B,2001ApJ...554..114E} 
\beq
c_{\mathrm{dm}} = 10\left(\frac{M_{cent}}{M_{\mathrm{nl}}(z)}\right)^{-0.13}
\eeq
where $M_{\mathrm{nl}}(z)$, the nonlinear mass scale, is defined such
that the rms linear density fluctuation extrapolated to redshift $z$
within a sphere containing mass $M_{\mathrm{nl}}$ is equal to
$\delta_c=1.686$, the linear overdensity at which a spherical
perturbation collapses.  $M_{\mathrm{nl}}(z)$ is a cosmology-dependent
parameter.  The satellites are assumed to be distributed according to
an NFW profile with concentration parameter $c_g = c_{\mathrm{dm}}$
\citep{1997ApJ...478..462C,2000MNRAS.318.1144P,2004MNRAS.352L...1G},
though the importance of this assumption -- which may overestimate $c_g$
\citep{1997ApJ...478..462C,2004ApJ...610..745L,2005ApJ...633..122H,2005ApJ...618..557N}
-- will be explored later. Unfortunately, since 
$c_g$ is, in general, poorly determined from previous data and our own,
we cannot fit for it and must assume some model.

The
conditional halo mass probability distribution for a given lens
luminosity $L_i$ is modeled as having
two parts, $p^C$ (central) and $p^{NC}$ (non-central), using a free
parameter $\alpha$, the satellite fraction:
\beq
p(M;L_i) = (1-\alpha) p^{C}(M;L_i) + \alpha\, p^{NC}(M;L_i).
\eeq
For halos hosting a 
central galaxy, we model the luminosity-halo mass relationship as a
delta-function; for halos hosting 
non-central galaxies, we assume that there is a
relationship between the number of galaxies of
this luminosity and the 
host halo mass, $\langle N\rangle(M;L) \propto M^{\epsilon}$
\citep[the host halo
number; ][]{1999MNRAS.303..188K,2004ApJ...609...35K,2004ApJ...610..745L,2005ApJ...630....1Z}.
As in \cite{2005MNRAS.362.1451M}, we use $\avgnm \propto M^{\epsilon}$ with 
$\epsilon=1$ for
$M>3M_{cent}$ and $\epsilon=2$ below that value.  This result was
shown there to match the simulations from \cite{2004ApJ...614..533T}
quite well, and the power-law exponent $\epsilon =1$ above a cut-off
is consistent with semi-analytic models of galaxy formation 
\citep{1999MNRAS.303..188K,2002MNRAS.335..311G}, N-body simulations
\citep{2004ApJ...609...35K,2005ApJ...630....1Z}, and observational measurements
\citep{1998ApJ...494....1J,2004ApJ...610..745L,2005astro.ph..9033C,2005ApJ...630....1Z}.
Since some of the observational analyses \citep[e.g.][]{2005ApJ...630....1Z} suggest that
significantly different values of $\epsilon$ may still be allowed, we 
explore in this paper the dependence of our results on its assumed value. 

Finally, our model for tidal stripping is that satellites have half of
their mass stripped, which corresponds to truncating the central density
profile at $0.4r_{vir}$ (beyond which we use $\ds \propto R^{-2}$ as for a
point mass). This assumption is consistent with the average mass loss for subhalos
observed in cosmological simulations \citep{2004MNRAS.355..819G,2005ApJ...618..557N}. We will
attempt to say something about this assumption by looking at galaxies
in low-density and high-density regions separately.

After using these inputs to obtain the galaxy-DM cross-power spectrum
$P_{g,dm}(k)$, we can Fourier transform to obtain the correlation
function $\xi_{g,dm}(R)$, integrate once
to get $\Sigma(R)$ via Eq.~\ref{E:sigmar}, and integrate again to
obtain $\ds(R)$ via Eq.~\ref{E:ds}.

Using the precomputed signal, we compare against
the lensing signal to derive two properties of the lens sample: the
halo mass $M_{cent}$ of the central galaxy, defined in terms of the radius
such that the overdensity within is equal to 180 times
the mean density (roughly 30
per cent larger than the mass $M_{200}$ defined within the radius where the
overdensity is equal to 200 times critical density, another definition 
of virial radius and mass that is 
often used); and the fraction of
galaxies in the sample that are satellites residing in a larger halo
(i.e., a group or cluster), $\alpha$.  Note that we do not attempt to
measure any other parameters of the central halo mass distribution
besides the central halo mass. In these fits, the average redshift of
the sample is relatively unimportant compared to the cosmology
($\Omega_m$ and $\sigma_8$), for which variation of best-fit satellite
fraction may be significant \citep{2005PhRvD..71d3511S}.

One issue raised in \cite{2005MNRAS.362.1451M} is the question of the
meaning of the best-fit halo mass; for samples of lenses with broad
central halo mass distributions, such that the mean and median are
significantly different, the best-fit masses must be adjusted upwards
if one is interested in the mean (as we are), and downwards if one is
interested in the median.  We attempt to isolate samples of lenses
with narrow distributions in central halo mass in two ways: first, by
using bins narrow in luminosity (half or one absolute magnitude or
a factor of 1.6-2.5
in luminosity) or stellar mass (a factor of 2 wide), and
second, by splitting the sample within those bins based on morphology
into early versus late types. Consequently, we use the small corrections to the
best-fit masses determined from Table 1 in \cite{2005MNRAS.362.1451M}
for the case of no scatter in the $M(L)$ relationship.  Details
of these corrections will be given in \S\ref{S:fitresults}.

\section{Data}\label{S:data}

The data used here are obtained from the SDSS
\citep{2000AJ....120.1579Y}, an ongoing survey to image roughly
$\pi$ steradians of the sky, and follow up approximately one million of
the detected objects spectroscopically \citep{2001AJ....122.2267E,
2002AJ....123.2945R,2002AJ....124.1810S}. The imaging is carried out by drift-scanning the sky
in photometric conditions \citep{2001AJ....122.2129H,
2004AN....325..583I}, in five bands ($ugriz$) \citep{1996AJ....111.1748F,
2002AJ....123.2121S} using a specially-designed wide-field camera
\citep{1998AJ....116.3040G}. These imaging data are the source of the
Large-Scale Structure (LSS)
sample that we use in this paper. In addition, objects are targeted for
spectroscopy using these data \citep{2003AJ....125.2276B} and are observed
with a 640-fiber spectrograph on the same telescope
\citep{telescope}. All of these data are 
processed by completely automated pipelines that detect and measure
photometric properties of objects, and astrometrically calibrate the data
\citep{2001adass..10..269L, 2003AJ....125.1559P,mtpipeline}. The SDSS is well
underway, and has had five major data releases \citep{2002AJ....123..485S,
2003AJ....126.2081A, 2004AJ....128..502A, 2005AJ....129.1755A,
2004AJ....128.2577F, 2005astro.ph..7711A}. 

\subsection{Lenses}

The galaxies used as lenses are those targeted as the DR4 MAIN
spectroscopic sample (4783 deg$^2$), with calibration as described in
\citet{2004AJ....128.2577F}.  Here we describe the quantities used for
classifying the lenses into 
subsamples: morphology, stellar masses, luminosities, and
density. We note that all samples described here are flux-limited, not
volume-limited, with the flux limit nominally $r<17.77$ (Petrosian,
extinction-corrected) but actually varying slightly in a known way
across the survey area.  Only those lens galaxies with redshifts
$0.02<z<0.35$ were used.

An important feature of this work is the splitting of lens samples
based on morphology within each stellar mass or luminosity subsample,
since predictions of the quantities we are measuring may differ
significantly for spiral and elliptical   
galaxies.  This split is carried out in practice by requiring that the
parameter 
{\tt frac\_deV} output from the {\sc Photo} pipeline be $\ge 0.5$ for
early types, $<0.5$ for late types.  This parameter
is determined by fitting the galaxy profile to {\tt frac\_deV} times
the best-fit deVaucouleurs profile plus $(1-${\tt frac\_deV}) times
the best-fit exponential profile (and requiring $0\le ${\tt
  frac\_deV}$\le 1$) in each band separately.  For this work, to
reduce the noise, we use the unweighted average over the $g$, $r$, and $i$
bands.  In practice, {\tt frac\_deV} is highly correlated with galaxy
colors.  \cite{2001AJ....122.1861S} studied the effectiveness of
morphological classification using either colors or de Vaucouleurs and
exponential profile likelihoods for a sample of spectroscopic galaxies
for which morphological classification was determined using 
spectral indices.  The result is that for early types, selection by
$r$-band {\tt frac\_deV} results in 96 per cent completeness, and is
76 per cent reliable (that is, 96 per cent of the early types
were indeed classified as early types, and 76 per cent of those
classified as early types were actually early types).  For late
types, the completeness was 55 per cent, and the classification was 90
per cent reliable.  However, since we use the average of the results
in $g$, $r$, and $i$, rather than just the results in $r$, and a more
recent version of {\sc Photo}, it is likely that the results of
classification using {\tt frac\_deV} for this paper are even better
than those quoted there.
We note that selection by $u-r$ color yielded 98 per cent completeness
and 83 per cent reliability for early types, and 72 per cent
completeness and 96 per cent reliability for late types, a slight
improvement in the results over classification using profiles in that
work, though our use of profile information from three bands and a
more recent version of {\sc Photo} may decrease that advantage.

Stellar mass estimates were obtained from the spectra using the same
techniques as in \cite{2003MNRAS.341...33K}, but for the full DR4
sample. A  library of 32,000 model star formation histories generated
using the \cite{2003MNRAS.344.1000B} population synthesis models and
the measured $D_n(4000)$ and $H\delta_A$ indices
\citep{2004ApJ...613..898T} are used to obtain a median likelihood
estimate of  the $z$-band mass-to-light ratio for each galaxy.  By
comparing the colour predicted by the  best-fit model to the observed
colour of the galaxy, the attenuation of the starlight due to dust is
also estimated. The corrected $z$-band magnitude and the $M/L_z$
together yield an estimate of the stellar mass. These estimates have
95 percent confidence ranges of around $\pm 40$ per cent. 
Comparison of these measures of stellar mass,
which are better measures of the mass in stars than luminosity,
against the best-fit halo masses will allow us to learn
about the efficiency of conversion of baryons in the halo to
stars. We define this efficiency via the relation
\begin{equation}\label{E:sf}
\eta = \frac{M_{stellar}}{M_{cent}}\frac{\Omega_m}{\Omega_b} = \frac{M_{stellar}}{M_{cent}f_b}
\end{equation}
where $f_b$ is the cosmological baryon fraction.
We note that this quantity is not the same as the traditional
definition of star formation efficiency, the rate at which stars are
forming compared to the available mass or surface mass density in HI
or molecular gas.  Hence, we use the term conversion efficiency rather
than star formation efficiency when describing these results.  There
are two differences between these quantities.  First, the star
formation efficiency denotes the fraction of those baryons that have
cooled off and formed part of the galaxy that then were included in
stars.  On the other hand, as shown in Eq.~\ref{E:sf}, the conversion
efficiency assumes that the baryon fraction in the galaxy is equal to
the cosmological one, thus ignoring the fact that some significant
fraction of baryons actually are not in galaxies, but are located in,
e.g., the hot gas in clusters, or warm-hot gas on the outskirts of
galaxies.  Furthermore, Eq.~\ref{E:sf} includes all baryonic matter,
such as helium, whereas the star formation efficiency is defined
relative to mass of hydrogen only.  Because of these differences in
definitions, our results for baryon conversion efficiency are also
lower limits on the star formation efficiency. 

Table~\ref{T:smsplit} gives  
information about the stellar mass bins used for this work, each of
which is a factor of two wide: the
numbers of galaxies in each bin, the mean redshift
within each bin (determined using the weights from the lensing
analysis), and the fraction of galaxies classified as spirals 
($f_{spiral}$).  We note that because of the
statistical error on each estimate, the bins are actually equivalent
to tophats a factor of two wide in stellar mass convolved with approximate
Gaussians with
$\sigma/\langle M_{stellar}\rangle \sim 0.2$.  Throughout this work,
stellar masses $M_{stellar}$ are given in units of $M_{\sun}$, having
been computed with $h=0.7$ as in \cite{2003MNRAS.341...33K}.  
\begin{table}
\caption{\label{T:smsplit}The stellar mass subsamples used in this
  analysis, including mean weighted redshifts within each
  bin, the total number of galaxies, and the fraction of spirals.}
\begin{tabular}{lcrrcr}
\hline\hline
Sample & $M_{stellar}$ & $N_{gal}$ & $\langle z \rangle$ &
$\langle M_{stellar}\rangle$ & $f_{spiral}$ \\
 & $[10^{10} M_{\sun}]$ & & & $[10^{10} M_{\sun}]$ & \\
\hline
sm1 & $[0.5, 1.0]$   & 23~474 & 0.060 & 0.74 & 0.74 \\
sm2 & $[1.0, 2.0]$   & 40~952 & 0.070 & 1.5 & 0.60 \\
sm3 & $[2.0, 4.0]$   & 66~503 & 0.085 & 2.9 & 0.46 \\
sm4 & $[4.0, 8.0]$   & 90~019 & 0.11 & 5.7 & 0.32 \\
sm5 & $[8.0, 16.0]$  & 82~734 & 0.13 & 11.0 & 0.20 \\
sm6 & $[16.0, 32.0]$ & 39~729 & 0.16 & 21.0 & 0.11 \\
sm7 & $[32.0, 64.0]$ &  8~096 & 0.19 & 40.0 & 0.05 \\
\hline\hline
\end{tabular}
\end{table}

We also must consider systematic uncertainties in our results due to
assumptions that are involved in the stellar mass
determination.  A full discussion can be found in
\cite{2003MNRAS.341...33K}; the main uncertainty that will affect our
results is the uncertainty due to the IMF, since we must assume that
the locally observed IMF applies universally.  Changing from
the \cite{2001MNRAS.322..231K} IMF to another IMF (assuming
that it is uniform across all samples), such as the Salpeter IMF, may
rescale all the 
stellar mass values, and therefore the derived conversion
efficiencies, by a fixed value of up to 30 per cent.  However, we note
that \cite{2005astro.ph..5042C} found with a sample of $>L_*$
ellipticals that the measured $M/L$
values appear inconsistent with the Salpeter IMF, which predicts too
much mass, so this value of 30 per cent is likely conservative.  
To address the question of whether the IMF really is uniform across
all samples, we note that \cite{2001MNRAS.322..231K} found no evidence
for significant variations in the IMF within our own galaxy; assuming
that the \cite{2005astro.ph..5042C} results represent typical
elliptical galaxies, and the MW is a typical spiral, we can then infer
that the IMF of spirals and 
ellipticals is not significantly different.
Thus, even if the global IMF leads to some overall rescaling of the
stellar mass values,
many of the results which we will present are in the form of trends of
conversion efficiency with stellar mass, or comparisons between
two different lens samples at fixed stellar mass, which are still
valid regardless of this systematic uncertainty.

Luminosities were determined using the $r$-band Petrosian apparent
magnitudes (extinction-corrected using the reddening maps from
\citealt{1998ApJ...500..525S}), k-corrections to $z=0.1$ (the sample
median) from {\sc kcorrect v1\_11} \citep{2003AJ....125.2348B}, passive
luminosity evolution  
correction from \citet{2003AJ....125.2276B}, and the distance modulus
determined using $h=1$, yielding $M_r = m + 1.6(z-0.1) - (K + DM)$.
These luminosities may be underestimated at typically the 10--20 per
cent level \citep{awest} due to the sky subtraction systematic
\citep{2005astro.ph..7711A,2005MNRAS.361.1287M}, an overestimate in the local
sky estimate within $\sim 90$'' of bright ($r < 18$) galaxies and
stars, thus affecting the entire lens sample at some level.
The actual underestimation varies with apparent magnitude and
size, and changes the
estimated $M/L$; since the effect has not been well-quantified for
the full sample and is
within the 95 per cent CL, we have
applied no correction.

One possible systematic in the comparisons between stellar mass and
luminosity is the fact that the luminosity is the full Petrosian
magnitude (which contains essentially all the light of an exponential
profile, and about 80 per cent of the light in a de Vaucouleurs
profile), whereas the stellar mass is estimated using spectra measured
using a 3'' aperture.  \cite{2003MNRAS.341...33K} discuss this
``aperture bias'' in detail, and while it is found to influence the
results at some level, the variation in $M_{stellar}/L$ due to this
systematic are insignificant compared to its variation with galaxy
properties such as luminosity and profile type (parametrized there
using the concentration index).

In order to obtain mass-to-light ratios, solar luminosities 
were determined using results from \citet{2003AJ....125.2276B} with
$M_* = -20.44$ and 
$M_{solar} = 4.76$, yielding luminosities in $h^{-2}L_{\sun}$ ($L* =
1.2h^{-2}\times 10^{10}L_{\sun}$).
Relevant information for luminosity subsamples is shown in
Table~\ref{T:lumsplit}, including numbers of galaxies, mean weighted
redshifts and luminosities, and spiral fractions.  We note that we
split the two brightest bins into half-magnitude bins, because we would like the
opportunity to better constrain the variation of mass with light in
that regime, and because the central halo mass distribution may become
wider at higher luminosities since some of the galaxies are
Brightest Cluster Galaxies (BCGs) of
clusters with large halo masses, and others are field galaxies hosted by 
smaller halos, giving
a broader halo mass distribution at the high-luminosity end.

\begin{table}
\caption{\label{T:lumsplit}The luminosity subsamples used in this
  analysis, including mean weighted redshifts and luminosities within each
  bin, the total number of galaxies, and the fraction of spirals.}
\begin{tabular}{lcrrcr}
\hline\hline
Sample$\!\!\!\!\!\!$ & $M_r$ & $N_{gal}$ & $\langle z \rangle$ &
$\langle L/L_*\rangle$ & $\!\!f_{spiral}\!\!$ \\
\hline
L1 & $-17 \ge M_r > -18$    &  10~047 & 0.032 & $\!\!$0.075$\!\!$ & 0.80 \\ 
L2 & $-18 \ge M_r > -19$    &  29~730 & 0.047 & 0.19  & 0.69 \\ 
L3 & $-19 \ge M_r > -20$    &  85~766 & 0.071 & 0.46  & 0.53 \\ 
L4 & $-20 \ge M_r > -21$    & $\!\!$141~976 & 0.10  & 1.1   & 0.35 \\ 
L5f & $-21 \ge M_r > -21.5$ &  60~994 & 0.14  & 2.1   & 0.23 \\ 
L5b & $-21.5 \ge M_r > -22$ &  34~920 & 0.17  & 3.2   & 0.16 \\ 
L6f & $-22 \ge M_r > -22.5$ &  13~067 & 0.20  & 4.9   & 0.08 \\ 
L6b & $-22.5 \ge M_r > -23$ &   2~933 & 0.22  & 7.7   & 0.05 \\
\hline\hline
\end{tabular}
\end{table}

Finally, we need a measure of the local galaxy environment.  
Many estimators, including nth nearest neighbor (3-d or in
projection), counts in an aperture (again, 3-d or in projection), and
Voronoi volumes, have been used in the
literature \citep[for
example,][]{2001A&A...368..776R,2002ApJ...580..122M,2003ApJ...585L...5H,2004MNRAS.348.1355B,2005ApJ...629..143B,2005ApJ...634..833C}.
Here, we choose a very simple 
one, spectroscopic galaxy counts in cylinders of radius 1
$h^{-1}$Mpc and line-of-sight length $\Delta v = \pm 1200$
km~s$^{-1}$.  These numbers are compared to
the number of random points in the same cylinders, thus
taking into account the angular and radial
variations in number density, survey boundaries, and other issues.
While galaxies excluded in the spectroscopic survey because of
fiber collisions are not used for the stellar mass
study (since they lack spectra), they are used for the
density estimates in order to avoid underestimating the environment
measure in rich clusters.  The galaxies without spectra due to fiber
collisions are given redshifts equal to those of the nearest
neighbor.  We note that
because the environment measurement requires a careful knowledge of
the survey completeness as a function of position, only those galaxies
included in the area covered by the LSS DR4 sample from the NYU
Value-Added Galaxy Catalog 
(VAGC; \citealt{2005AJ....129.2562B}) were used to obtain
environment estimates.  

Our estimates were derived using 20 times as many random points as
real galaxies; we note that for higher redshifts, these environment
estimates can be quite noisy or may fail entirely, because the radial
selection function 
leads to a low density of objects.  Furthermore, galaxies at the
lower redshift
limit or near survey boundaries may not have environment estimates if no random
galaxies were found in the cylinders around them; thus, when splitting
the sample at the median density within each stellar mass or
luminosity bin, only 
some fraction of the sample was used (ranging from 93 per cent for
bins at $L <\sim L_*$, down to 40 per cent for the
brightest galaxies).  Fortunately, the lensing signal itself provides
a reasonable test of the environment estimate, because we can see
whether the signal at 1--2 $h^{-1}$Mpc scales is consistent with the
lens sample primarily being in the field or in groups and clusters.
At fixed luminosity or stellar mass, the spiral 
sample (defined by {\tt frac\_deV} $< 0.5$) had a lower median and mean
environment estimate than ellipticals, as expected from previous
studies cited in \S\ref{S:intro}.

Because of the need for spectra to determine stellar masses, our
sample must exclude galaxies for which spectra were not obtained due
to fiber collisions.  Fibers cannot be placed closer than 55''
($\sim 80$ \hinvk\ at $z\sim 0.1$), so
if two targets are closer than this separation, only one will have a
spectrum.  This restriction is alleviated in roughly one third of the
sky by the use of overlapping plates, but nonetheless, roughly 7 per
cent of targets do not have spectra.  This loss of targets is
naturally worse in high-density regions, and will therefore tend to
affect the satellite contribution, decreasing it  in a
scale-dependent way, and therefore
changing its shape.  However, as our
results will show, we are not highly sensitive to the shape of the
satellite contribution.

\subsection{Sources}

The source sample is the same as that from \citealt{2005MNRAS.361.1287M}
(hereinafter M05), who
obtained shapes for more than 30 million galaxies in the SDSS imaging data
down to magnitude $r=21.8$ (i.e. four magnitudes fainter than the SDSS
spectroscopic limit).  This section briefly describes the M05
pipeline, also known as Reglens. 

The M05 pipeline obtains galaxy images in the $r$
and $i$ filters from the SDSS ``atlas images''
\citep{2002AJ....123..485S}.  The basic principle of shear measurement
using these images is to fit a Gaussian profile with elliptical isophotes
to the image, and define the components of the ellipticity
\beq
(e_+,e_\times) = \frac{1-(b/a)^2}{1+(b/a)^2}(\cos 2\phi, \sin 2\phi),
\label{eq:e}
\eeq
where $b/a$ is the axis ratio and $\phi$ is the position angle of the
major axis.  This is then an estimator for the shear,
\beq
(\gamma_+,\gamma_\times) = \frac{1}{2\cal R}
\langle(e_+,e_\times)\rangle,
\eeq
where ${\cal R}\approx 0.87$ is called the ``shear responsivity'' and
represents the response of the ellipticity (Eq.~\ref{eq:e}) to a small
shear \citep{1995ApJ...449..460K, 2002AJ....123..583B}.  In practice, a
number of corrections need to be applied to obtain the ellipticity.  The
most important of these is the correction for the smearing and
circularization of the galactic images by the PSF;
M05 uses the PSF maps obtained from stellar images
by the {\sc psp} pipeline \citep{2001adass..10..269L}, and corrects
for these
using the re-Gaussianization technique of \citet{2003MNRAS.343..459H}.
Re-Gaussianization corrects for the PSF while taking into account the
non-Gaussianity both of the PSF and of the galaxy profile.  A
smaller correction is for the optical distortions in the telescope:
ideally the mapping from the sky to the CCD is shape-preserving
(conformal), but in reality this is not the case, resulting in a nonzero
``camera shear.'' In the SDSS, this is a small effect (of order 0.1
per cent) which can be identified and removed using the astrometric solution
\citep{2003AJ....125.1559P}.  Finally, a variety of systematics tests are
necessary to determine that the shear responsivity ${\cal R}$ has in fact
been determined correctly.  We refer the interested reader to
M05 for the details of these corrections and
tests.

M05 includes a
lengthy discussion of shear calibration biases in this catalog;
we will only summarize these issues briefly here.  Our source sample
is divided into three subsamples: $r<21$, $r>21$, and high-redshift
Luminous Red Galaxies (LRGs, \citealt{2001AJ....122.2267E}), defined
using color and magnitude cuts as in M05 using selection 
criteria related to those from \cite{2001AJ....122.2267E} and
\cite{2005MNRAS.359..237P}.  Using simulations 
from \citet{2003MNRAS.343..459H} to estimate the PSF dilution
correction and analytical models for selection biases and other issues
that affect shear calibration, we place $2\sigma$ limits on the
multiplicative shear
calibration bias of $[-5, 12]$ per cent for $r<21$, $[-8, 18]$ per
cent for
$r>21$, and $[-6, 19]$ per cent for LRGs.

As shown in Eq.~\ref{E:ds}, the lensing signal $\ds$ is a product of
the shear and factors involving lens and source redshifts.  Since the
lenses have spectroscopic redshifts, the primary difficulty is
determining the source redshift distribution.  We take three
approaches, all described in detail in M05.
For the $r<21$ sources, we use photometric redshifts and their error
distributions determined using a sample of galaxies in the Groth strip
with redshifts from DEEP2 \citep{2003SPIE.4834..161D,2003ApJ...599..997M,2004ApJ...609..525C,2004astro.ph..8344D}, and require
$z_s>z_l+0.1$ to avoid contamination from physically-associated
lens-source pairs.  For the $r>21$ sources, we use redshift
distributions from DEEP2.  For the high-redshift LRGs, we use
photometric redshifts and their error distributions determined using
data from the 2dF-Sloan LRG and Quasar Survey (2SLAQ), and presented in
\cite{2005MNRAS.359..237P}.

Finally, we have placed constraints on other issues affecting the
calibration of the lensing signal, such as the sky subtraction
problem, intrinsic alignments, magnification bias, star-galaxy
separation, and seeing-dependent systematics.  As shown in
M05 the calibration of the signal using the
three source samples agrees to within 10 per cent, with a total
$1\sigma$ calibration uncertainty estimated at 
7 per cent ($r<21$) or 10 per cent ($r>21$ and LRG).

\subsection{Signal computation}

Here we describe the computation of the lensing signal.  Lens-source
pairs are assigned weights according to the error on the shape
measurement via
\beq
w_{ls} = \frac{\Sigma_c^{-2}}{\sigma_s^2 + \sigma_{SN}^2}
\eeq
where $\sigma_{SN}^2$, the intrinsic shape noise, was determined as a
function of magnitude in M05, figure 3.  The factor of
$\Sigma_c^{-2}$ downweights pairs that are close in redshift.

Once we have computed these weights, we compute the lensing signal in
46 logarithmic radial bins from 20 $h^{-1}$kpc to 2 $h^{-1}$Mpc as
a summation over lens-source pairs via:
\beq
\ds(R) = \frac{\sum_{ls} w_{ls} \gamma_t^{(ls)} \Sigma_c}{2 {\cal
    R}\sum_{ls} w_{ls}} 
\eeq
where the factor of 2 arises due to our definition of ellipticity.

There are several additional procedures that must be done when
computing the signal (for more detail, see M05).  First, the signal
computed around random points must be subtracted from the signal
around real lenses to eliminate contributions from systematic shear.
Second, the signal must be boosted, i.e. multiplied by $B(r) =
n(r)/n_{rand}(r)$, the ratio of the number density of sources relative
to the number around random points, in order to account for dilution
by sources that are physically associated with lenses, and therefore
not lensed.

In order to determine errors on the lensing signal, we divide the
survey area into 200 bootstrap subregions, and generate 2500
bootstrap-resampled datasets.  To determine errors on fit parameters,
we perform the fits on each bootstrap-resampled dataset, and use the
distribution in parameter space to determine confidence intervals.

\section{Results}\label{S:results}
\subsection{Lensing signal}

\begin{figure*}
\includegraphics[width=6.5in]{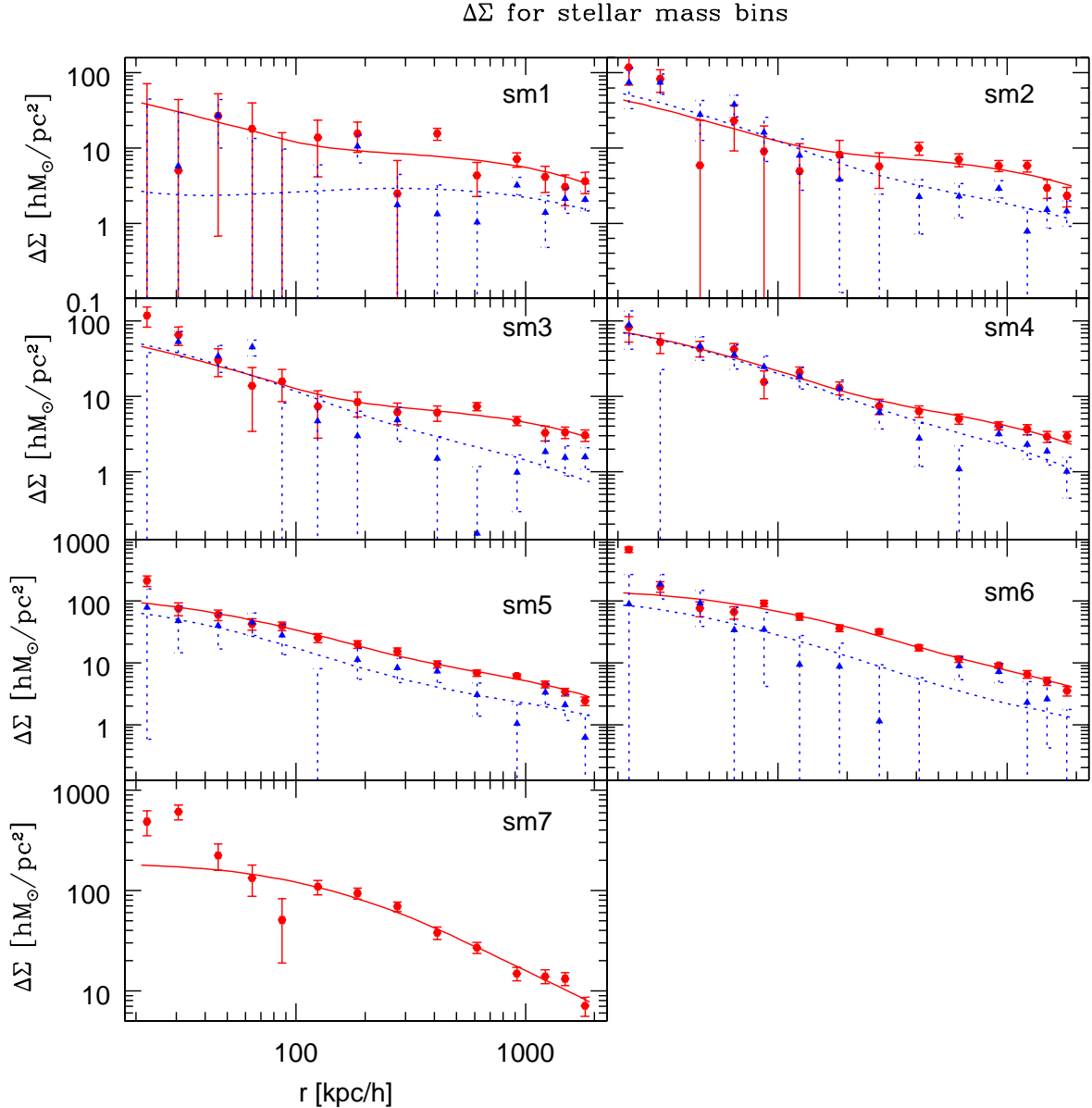}
\caption{\label{F:small}\ds{} in stellar mass bins for early (red
  hexagons, solid line)
  and late-type (blue triangles, dashed line) galaxies.  For the highest stellar mass bin,
  only the signal for early-types is shown since they are 95 per cent
  of the sample.  All errors are $1\sigma$.} 
\end{figure*}

\begin{figure*}
\includegraphics[width=6.5in]{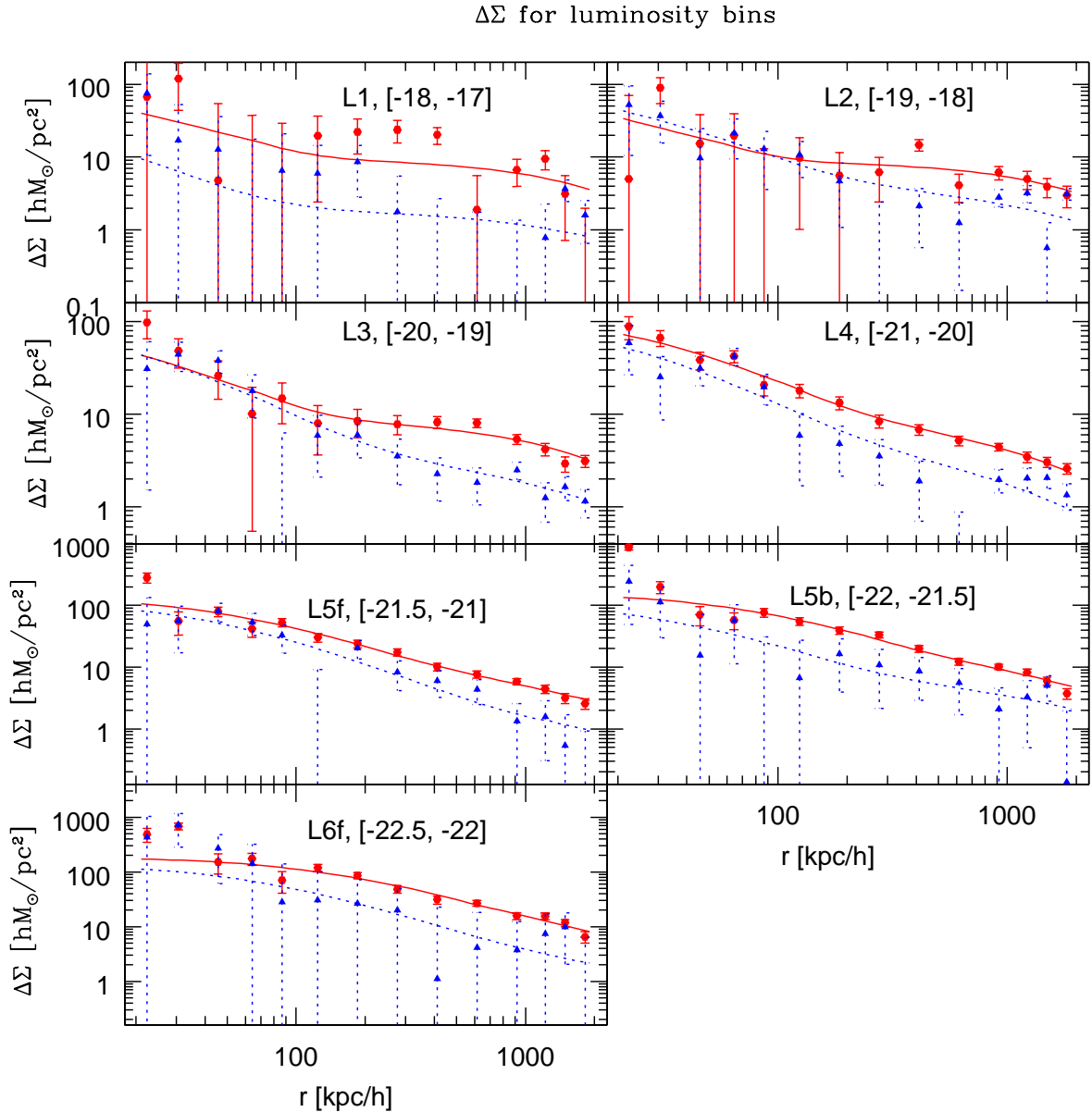}
\caption{\label{F:lumall}\ds{} with best-fit halo model in luminosity
  bins for early (red hexagons, solid line)
  and late-type (blue triangles, dashed line) galaxies.  
All
errors are $1\sigma$.} 
\end{figure*}

\begin{figure*}
\includegraphics[width=6.5in]{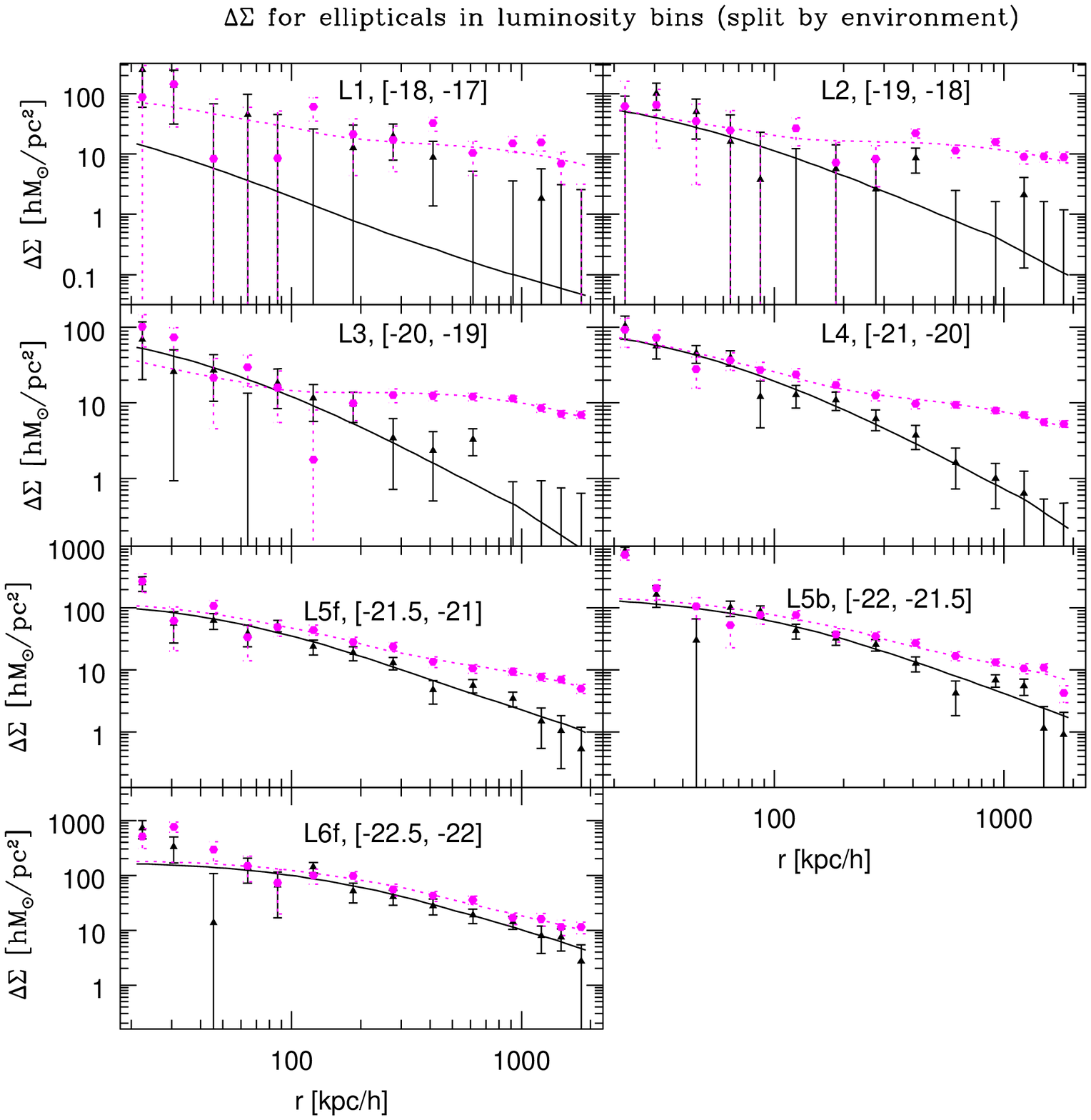}
\caption{\label{F:lumdenall}\ds{} with best-fit halo model in
  luminosity bins for early-type 
  galaxies, divided into overdense (magenta hexagons, dashed line) and
  underdense (black triangles,
  solid line) samples.  All errors are $1\sigma$.} 
\end{figure*}

In Fig.~\ref{F:small} we show the lensing signal in stellar mass bins
with $1\sigma$ errors,
for spiral ({\tt frac\_deV}$<0.5$) and elliptical ({\tt frac\_deV}$\ge
0.5$) galaxies, with the best-fit halo model signal (with 
parameters to be described in \S\ref{S:fitresults}).  Fig.~\ref{F:lumall}
shows the same for luminosity bins; the signal was noisy enough for
the brightest bin (L6b) that errors could not be determined on fits,
so that bin is not shown.  Finally, Fig.~\ref{F:lumdenall} shows the
results in luminosity bins for low- and high-density samples of
ellipticals only.  The
results have been rebinned for easier viewing; correlations between
radial bins are minimal at small radius, and reach a maximum of about
0.2--0.3 at the outermost bins for the lower luminosity or stellar
mass bins (at lower $z$) and are negligible at all radii for the
brighter bins.  While the model fits are shown on the plots, we defer
discussion of the best-fit parameters to the following sections, and
focus here on a comparison of the lensing signal.

First, we consider Fig.~\ref{F:small}.  There appears to be some
detection of signal on all scales for all but the lowest bin, $5\times 10^9 <
M_{stellar}/M_{\sun} < 10^{10}$.  For sm2--sm4, comprising
$(1-8)\times 10^{10} M_{\sun}$, the signal for early and late
types appears statistically consistent for $r<\sim 100$ \hkpc,
indicating that the average halo masses (and therefore conversion
efficiencies) for the two types of galaxies
are similar.  For sm5 and sm6, comprising $(8-32) \times 10^{10}
M_{\sun}$, the signal on small scales is larger for early types than
for late types, indicating a larger average halo mass for early types,
so stellar mass is no longer a good tracer of halo mass for this range
of stellar masses.  This result is not entirely surprising, since for
these bins, some fraction of the red galaxies are LRGs that may be the
BCGs of clusters, so the halo masses determined using
lensing are actually the masses of the entire cluster rather than of
individual galaxies.  On larger scales, the signal for early types
is higher than 
for late types for each stellar mass bin to a degree that is highly
statistically significant, consistent with previous
findings that red galaxies reside in denser environments than blue
galaxies.  

%
Next, we consider Fig.~\ref{F:lumall}.  The detection is marginal in
L1 and L2, comprising magnitudes $-19 \le M_r < -17$.  For the bins with a
robust detection, L3 (with $\langle L/L_*\rangle=0.46$) is the
only one for which the signal for early 
and late types appears to be highly consistent at small scales; at
higher luminosity, it appears to be higher for early types.  On large
scales, the signal 
for early types is again larger than for late types for all luminosity
bins, indicating a
tendency for early types to be in denser environments.  

Finally, we consider Fig.~\ref{F:lumdenall}, which shows \ds{} in
luminosity bins for early-type galaxies only, with the sample divided
at the median environment measure in that bin.  We remind the reader
that environment estimates were not obtained for a large fraction of
the galaxies in the brightest bins due to the limitations of the
environment estimator, so few lenses were used for those bins.  For
L1--L5, it is 
apparent from the signal on $r > 300$ \hkpc{} that the environment
split efficiently separates 
the galaxies into those that reside in overdense versus underdense
regions.  This separation is particularly marked for L3 and L4, with
the signal on 1--2 \hmpc{} scales differing between the high- and
low-density samples by more than an
order of magnitude.  For the brightest bins, the difference is less
striking; it is possible that our environment estimator is simply too
noisy at higher redshifts, or that enough of the L6 early-type
galaxies reside in 
groups and clusters that the environment simply does not vary much
across the bin, leading to a small variation in signal.  For all bins,
the signal on small scales does 
not vary significantly with the environment. As we show below, the 
$L_*$ high density sample is dominated by satellites. We see 
that the tidal stripping of dark matter around satellites
inside groups and clusters
cannot be maximally efficient, since
this would have been seen as a suppression of signal on small scales.

\subsection{Halo model fits}\label{S:fitresults}

Here we discuss the goodness-of-fit for the halo model fits before
proceeding to discussions of the best-fit parameters.  However, we
note that the values of $\chi^2$ are not expected to follow the usual $\chi^2$
distribution because of the noisiness of bootstrap covariance
matrices (see \citealt{2004MNRAS.353..529H} for a fuller description of
these results). For our fits with 40 degrees of freedom and
200 bootstrap regions, the expected value of $\chi^2$ is 50.4, not
40.  It is clear from the figures that the halo model signal generally
is a close match to the data; for example, for early types, the
$\chi^2$ for the fits for the seven stellar mass bins were 37.5, 39.9,
31.5, 36.5, 46.9, 40.5, and 42.0
respectively, with $p$-values (i.e. the probability to exceed this
value by chance if the model is a realistic description of the data)
of 0.85, 0.79, 0.96, 0.87, 0.58, 0.78, 0.73.  The results may imply
that the bootstrap errorbars are slight overestimates. Results for
late types in stellar mass bins, and the
splits by luminosity, were similar.

One concern regarding these fits is that 
for the highest stellar mass and luminosity bins the halo model underestimates the
signal by a significant amount on small scales ($r<\sim 40$ \hkpc).
Because of the sky subtraction problem
\citep{2005astro.ph..7711A,2005MNRAS.361.1287M},  only scales $r>30$
\hkpc{} are used for the fits 
even though we have plotted down to 20 \hkpc{}, so this discrepancy
between the best-fit and observed signal on these scales does not
cause a major increase in the $\chi^2$; however, it 
is still a concern, since the magnitude of the effect is larger than
the estimated effect due to sky subtraction (at most 10 per cent).
Because the large magnitude of this discrepancy 
does not seem attributable to any of the small-scale systematics such
as intrinsic alignments, the sky subtraction problem, or magnification
bias, the discrepancy seems to suggest a problem with the assumed
profile itself, likely due to the 
effects of stellar component and the associated dark matter contraction 
caused by it, which we have completely ignored.  A detailed study of
this discrepancy  
 between the model and the observed signal will be investigated 
in the context of dark matter profile constraints from g-g lensing
in future work.

One final issue we must consider is that brought up in
\cite{2005MNRAS.362.1451M} of the corrections to the central halo
masses due to broadness of the central halo mass distribution, which
naturally has some width due to the width of the stellar mass or
luminosity bins, and which may be further broadened by scatter in the
mass-luminosity (or mass-stellar mass) 
relationship.  As noted there, the best-fit central halo mass
underestimates the mean value (and overestimates the median, which is
not of interest in this work) by some amount due to this width.  In
this work, we use the corrections in the ``no-scatter case'' which are
purely due to the width of the bins that lead to a wide central halo
mass distribution, because we anticipate that our division into early
and late type samples will lead to narrower central halo mass
distributions.  Those corrections, as described there, are to 
increase the best-fit masses by 6 per cent, 11 per cent, and 36 per
cent for L3, L4, and L5 (1 magnitude wide), respectively.  We
extrapolate these values to higher luminosity and use no correction
for L1--L2, 6 per cent, 11 per cent, 25 per cent, 50 per cent, and 66
per cent for L3, L4, L5f, L5b, and L6f, respectively.  Fundamentally
these values are somewhat uncertain, and indeed represent one of the
main modeling uncertainties of this paper. 
All corrections have been
applied to values in the tables and text in the following sections.
For stellar mass samples, we use the average luminosity of each bin to find a
correspondence between stellar mass and  luminosity bins, and use
corrections of 0, 6, 8, 
11, 22, 45, and 66 per cent for stellar mass bins 1--7, respectively.
While these corrections introduce some uncertainty in the
masses themselves, they do not affect comparisons between morphology
samples at the same stellar mass or luminosity since the same
corrections are applied to each morphology sample.

\subsection{Central halo masses}

Here we present results derived from the best-fit central halo masses
and their 95 
per cent confidence intervals.  Due to the imposition of the
constraint $M_{cent} > 0$, the error distributions at low luminosity and
stellar mass are highly non-Gaussian, and the confidence intervals are
determined using the distribution of $M_{cent}$ values for the bootstrap
subsamples.

\begin{figure}
\includegraphics[width=3.0in]{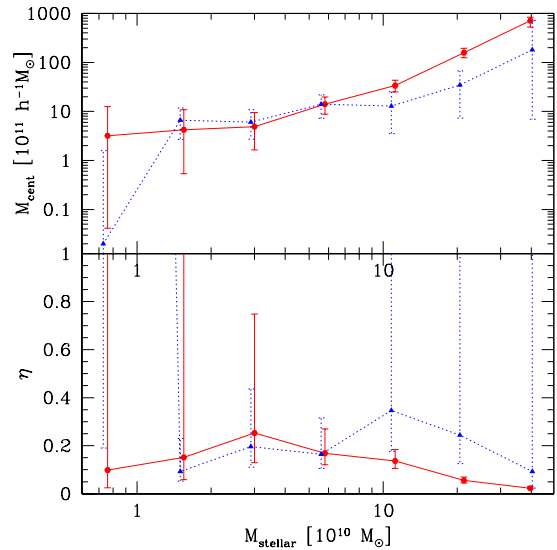}
\caption{\label{F:smtrends}Central halo mass (top) and conversion
  efficiency in central galaxy $\eta$ (bottom) as a
  function of stellar mass for early-type (red hexagons, solid line) and late-type
  (blue triangles, dotted line) galaxies.  All errorbars shown are 95 per cent confidence
  interval.}  
\end{figure}

First, we consider the relationship between halo mass and stellar
mass.  Table~\ref{T:massvssm} shows the best-fit halo mass and its 95 per cent
confidence interval as a function of stellar mass for early and late
type galaxies.  For reference, the average $r$-band luminosity for each
sample is also shown.  Figure~\ref{F:smtrends} shows a plot of halo mass as a
function of stellar mass, and of central galaxy conversion efficiency as a
function of stellar mass, where we define conversion efficiency $\eta$
via Eq.~\ref{E:sf}
with $\Omega_m=0.27$, $\Omega_b=0.046$, and $h=0.7$.

\begin{table}
\caption{\label{T:massvssm}The central halo mass determined for
  stellar mass subsamples separately for early and late type
  galaxies, and the resulting conversion efficiencies determined
  according to Eq.~\ref{E:sf}.  All confidence intervals are 95 per
  cent.}
\begin{tabular}{lccc}
\hline\hline
$\langle M_{stellar}\rangle$ & $M_{cent}$ & $\eta$ & $\langle
L/L_*\rangle$ \\
 $10^{10} M_{\sun}$ & $10^{11} h^{-1}M_{\sun}$ & & \\
\hline
\multicolumn{4}{c}{Early types}\\
\hline
0.76 & $3.18_{-3.14}^{+9.45}$ & $0.098_{-0.073}^{+7.44}$ & 0.27 \\
1.5 & $4.20_{-3.67}^{+6.63}$ & $0.15_{-0.09}^{+1.00}$ & 0.41 \\
3.0 & $4.9_{-3.2}^{+4.7}$ & $0.25_{-0.12}^{+0.50}$ & 0.67 \\
5.8 & $14.1_{-5.3}^{+5.6}$ & $0.17_{-0.05}^{+0.10}$ & 1.1 \\
11.2 & $34_{-9}^{+10}$ & $0.14_{-0.03}^{+0.05}$ & 1.8 \\
21.3 & $158_{-33}^{+37}$ & $0.055_{-0.010}^{+0.015}$ & 2.9 \\
39.6 & $716_{-190}^{+123}$ & $0.023_{-0.003}^{+0.008}$ & 4.9 \\
\hline
\multicolumn{4}{c}{Late types}\\
\hline
0.73 & $0.020_{-0.018}^{+1.56}$ & $15.0_{-14.8}^{+107}$ & 0.38 \\
1.5 & $6.6_{-4.0}^{+5.1}$ & $0.09_{-0.04}^{+0.14}$ & 0.63 \\
2.9 & $6.1_{-3.4}^{+4.7}$ & $0.20_{-0.09}^{+0.24}$ & 0.77 \\
5.6 & $14_{-7}^{+8}$ & $0.16_{-0.06}^{+0.15}$ & 1.1 \\
10.8 & $13_{-9}^{+12}$ & $0.35_{-0.17}^{+0.92}$ & 1.9 \\
20.5 & $34_{-28}^{+33}$ & $0.24_{-0.12}^{+0.90}$ & 3.2 \\
40.4 & $180_{-173}^{+532}$ & $0.09_{-0.07}^{+2.32}$ & 3.8 \\
\hline\hline
\end{tabular}
\end{table}

Table~\ref{T:massvssm} and Fig.~\ref{F:smtrends} reveal several
interesting trends in the relationship between stellar mass and halo
mass.  First, stellar mass is a good tracer of halo mass for
$M_{stellar} <~ 10^{11} M_{\sun}$ (corresponding roughly to $L <~
1.5L_*$); that is, both early- and 
late-types have the same average relationship between halo and stellar
mass in this regime. Above this value of stellar mass, the central
halo mass for early types is larger than that for late types.  One
possible explanation for this trend is that early types are far more
likely to be the central galaxy of a group- or cluster-mass halo than
late types, which are only at the center of smaller halos.
Thus, the $M_{cent}$ values in this table and figure for early types
at higher values of stellar mass are likely to reflect the
mass of the entire group/cluster rather than that of the central
galaxy only.  

We also consider the lower panel of Fig.~\ref{F:smtrends}, the
central galaxy 
conversion efficiency $\eta$ as a function of $M_{stellar}$.  For
those bins with a marginal detection of signal, we are able to place
only weak constraints on $\eta$, which may at times even have a central
value outside of the range of reasonable values $0<\eta<1$; however,
in all cases the $2\sigma$ errors shown in the table and figure
include part of the range of reasonable values.   The
particularly useful part of this exercise is the lower-limit in
$\eta$ derived from the upper-limit on the halo masses.  For
$M_{stellar} > 10^{11} M_{\sun}$, we see a reflection of the trend noted for
the early types in the upper panel, that many of them are
central galaxies of groups or clusters, and consequently the
interpretation of $M_{stellar}/M_{cent}$ in terms of conversion
efficiency is no longer reasonable.  (To do so, we would
have to include the stellar masses of all other group/cluster
members or include the mass of the central galaxy only rather than the
full cluster mass.)  However, including stellar mass of
other group/cluster members would tend to raise $\eta$, so while the
central value of $\eta$ is likely to be underestimated at high
$M_{stellar}$, the lower 
limits given here are robust.  Hence, for late type galaxies, we can
conclude that for typical galaxies with $M_{stellar} \sim 10^{11}
M_{\sun}$, the conversion efficiency is typically 35 per cent,
with a lower limit of 18 per cent (95 per cent CL).  For early types,
the typical galaxy has a conversion efficiency of only 14 per
cent, or a lower limit of 11 per cent.  These results imply a difference in
conversion efficiency between typical early and late type galaxies
of roughly a factor of two or more, with the efficiency for both early and
late types declining at higher stellar masses.

\begin{figure}
\includegraphics[width=3.0in]{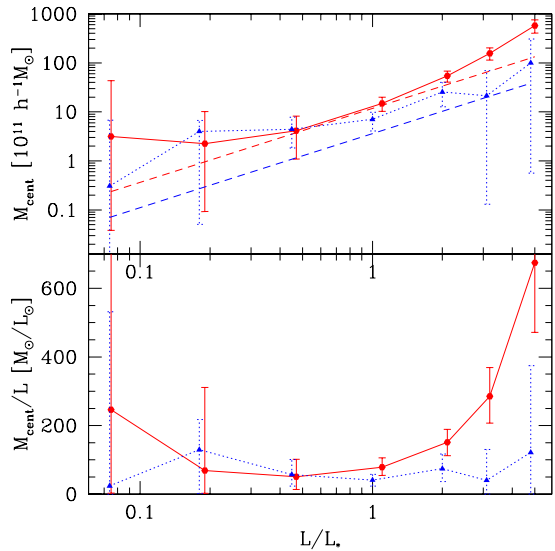}
\caption{\label{F:lumtrends}Central halo mass (top) and $M_{cent}/L$ as a
  function of luminosity for early-type (red hexagons, solid line) and late-type
  (blue triangles, dotted line) galaxies.  Dashed lines also show the best-fit power-law
  $M(L)$ for early and late types from \protect\cite{2002MNRAS.335..311G} as
  discussed in the text.  All errorbars shown are 95 per cent confidence
   intervals.}  
\end{figure}

Next, we consider the relationship between $r$-band luminosity and
halo mass.  Table~\ref{T:massvslum} and Fig.~\ref{F:lumtrends} give
the relevant parameters.  
\begin{table*}
\caption{\label{T:massvslum}The central halo mass determined for
  luminosity subsamples separately for early and late type
  galaxies.  All confidence intervals are 95 per
  cent.}
\begin{tabular}{cccccc}
\hline\hline
$M_r$ & $\langle L/L_*\rangle$ & $M_{cent}$ & $M_{cent}/L$ &
$M_{cent}$ & $M_{cent}/L$ \\
 & & $10^{11} h^{-1}M_{\sun}$ & $M_{\sun}/L_{\sun}$ & $10^{11} h^{-1}M_{\sun}$ & $M_{\sun}/L_{\sun}$ \\
\hline
 & & \multicolumn{2}{c}{Early types} & \multicolumn{2}{c}{Late types}\\
\hline
$[-18, -17)$ & 0.075 & $3.16_{-3.13}^{+40}$ & $246_{-243}^{+3113}$ & $0.31_{-0.30}^{+6.44}$ & $24_{-24}^{+507}$ \\
$[-19, -18)$ & 0.19 & $2.2_{-2.1}^{+7.9}$ & $69_{-66}^{+242}$ & $4.0_{-3.9}^{+2.7}$ & $129_{-128}^{+88}$ \\
$[-20, -19)$ & 0.47 & $4.1_{-3.0}^{+4.1}$ & $51_{-37}^{+51}$ & $4.4_{-2.6}^{+3.4}$ & $57_{-33}^{+44}$ \\
$[-21, -20)$ & 1.1 & $14.9_{-4.6}^{+5.0}$ & $79_{-24}^{+27}$ & $7.1_{-3.0}^{+2.8}$ & $41_{-17}^{+16}$ \\
$[-21.5, -21)$ & 2.1 & $54\pm 14$ & $151\pm 38$ & $26_{-13}^{+15}$ & $74_{-37}^{+43}$ \\
$[-22, -21.5)$ & 3.2 & $157_{-43}^{+46}$ & $285_{-78}^{+83}$ & $21_{-21}^{+48}$ & $40_{-40}^{+90}$ \\
$[-22.5, -22)$ & 5.0 & $578_{-174}^{+180}$ & $674_{-203}^{+210}$ & $100_{-100}^{+208}$ & $122_{-121}^{+253}$ \\
\hline\hline
\end{tabular}
\end{table*}
This table and figure reveal several interesting points about the
relationship between $r$-band luminosity and halo mass.  First, the
luminosity is a good tracer of central halo mass for $L<\sim 0.8L_*$,
with early- and late-types below this luminosity having consistent
halo masses.  At higher
luminosities, we see a similar trend as for the split by stellar
masses, with larger central halo masses at the bright end for early
types, possibly explained by them being at the centers of group- or
cluster-mass halos.  Hence, while the relationship between mass and
luminosity appears to be consistent with a single power-law for late
types, the same is not true for early types, due to the steepening in
this relationship at the bright end (which, in this regime, is more
accurately a relationship between the BCG luminosity and the cluster
mass rather than an individual galaxy luminosity and mass).  In the
lower panel of Fig.~\ref{F:lumtrends}, we note that because of this
inclusion of the mass in all group/cluster members, the central values
and lower limits of $M_{cent}/L$ are not as relevant as the upper
limits.  Consistent with the results derived using stellar masses,
$M_{cent}/L$ is larger for early types than for late types by about a
factor of two around $L_*$, and the value of $M_{cent}/L$ decreases
significantly for higher luminosities.  As for the relationship
between halo mass and stellar mass for isolated halos,
we need to use environment estimates to explore what happens to field
galaxies at high luminosities in order for the values for early types
to have a simple physical interpretation.

\begin{figure}
\includegraphics[width=3.0in]{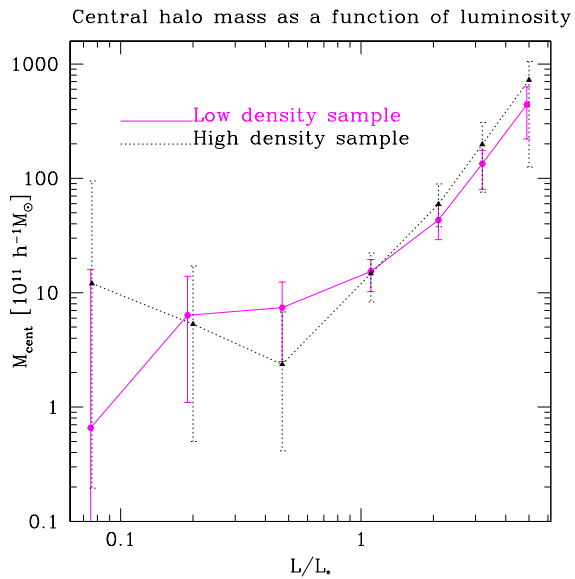}
\caption{\label{F:dentrends}Central halo mass as a
  function of luminosity for early-type galaxies in low-density
  (magenta hexagons, solid line) and
  high-density (black triangles, dotted line) samples.  All errorbars
  shown are 95 per cent confidence intervals.}  
\end{figure}

Next, we consider the relationship between local density and
halo mass for ellipticals, where we divide each luminosity bin at the
median density in that bin. 
Table~\ref{T:massden}  and 
Fig.~\ref{F:dentrends} contain the results.  
\begin{table}
\caption{\label{T:massden}The central halo mass determined for
  luminosity subsamples separately for early-type galaxies in low- and
  high-density regions.  All confidence intervals are 95 per
  cent.}
\begin{tabular}{lcc}
\hline\hline
$\langle L/L_*\rangle$ & $M_{cent}$ $[10^{11} h^{-1}M_{\sun}]$ & $M_{cent}$ $[10^{11} h^{-1}M_{\sun}]$\\
 & Low density & High density \\
\hline
0.075 & $0.66_{-0.66}^{+15}$ & $12.0_{-11.8}^{+82.8}$ \\
0.19 & $6.3_{-5.2}^{+7.6}$ & $5.3_{-4.8}^{+11.8}$ \\
0.47 & $7.4_{-4.9}^{+5.0}$ & $2.4_{-1.9}^{+4.4}$ \\
1.1 & $15.4_{-5.1}^{+4.1}$ & $15_{-6}^{+8}$ \\
2.1 & $43_{-14}^{+15}$ & $59_{-21}^{+30}$ \\
3.2 & $135_{-54}^{+41}$ & $198_{-122}^{+109}$ \\
4.9 & $443_{-222}^{+190}$ & $725_{-599}^{+329}$ \\
\hline\hline
\end{tabular}
\end{table}

The signal in Fig.~\ref{F:lumdenall}, and the best-fit masses in
Table~\ref{T:massden} and Fig.~\ref{F:dentrends} appear to suggest
that early-type galaxies in both low-density and high-density regions
have consistent central halo masses, particularly for $L<\sim L_*$
where the high-density sample is exclusively satellites, rather than
BCGs.  Another interesting point to note is that discussed in
\cite{2005MNRAS.362.1451M}, that a large amount of
tidal stripping of satellite profiles in clusters would cause a
clear depression in the lensing signal on scales below the virial
radius.  For example, a scenario predicted from a combination of
N-body simulations with semi-analytic models of galaxy formation 
suggested by \cite{2004MNRAS.352L...1G} is that a very large
fraction may have nearly all the dark matter stripped.  (As stated in
\S\ref{S:hm}, we do assume that all 
satellites have 50 per cent of the matter stripped.)  No such major
depression in the signal relative to that for field galaxies is seen
on $50$--$100$ \hkpc{} scales,
and therefore we can rule out scenarios that have  most satellites
strongly stripped. 

\subsection{Comparison against past studies}

Other works have attempted to determine whether halos of satellite
galaxies are tidally stripped.  The methodology presented by
\cite{1997MNRAS.287..833N} 
allowed for the mass distributions of individual cluster galaxies to
be mapped out, in addition to the large-scale cluster mass
distribution.  This method was applied by \cite{1998ApJ...499..600N},
who found some suggestion that tidal stripping may affect $E$ galaxies
less than $S0$ galaxies (based on differences in velocity
dispersions), 
and \citealt{2002ApJ...580L..11N} (with a longer description of
methodology in \citealt{2004astro.ph.11426N}), who found clear
evidence for tidal 
stripping, with field-scale halos for $L_*$ early types excluded at
the $10\sigma$ level.  Their density profile for the galaxy-scale
halos differed from the one used in this work, but shares the property
that below the truncation radius, $\Delta\Sigma \propto R^{-1}$ (our
data is consistent with a power-law index of $-0.9$, so quite
similar), and $\propto R^{-2}$ beyond truncation radius $r_t$, which
is the same as our model, but with the transition happening gradually
rather than abruptly as in our model.

For a typical cluster, the
truncation radius for 
$L_*$ early type galaxies was found in \cite{2002ApJ...580L..11N} to
be $40$ kpc, or $28$ \hkpc{}; 
our best-fit value of $M_{cent}=15_{-6}^{+8} \times 10^{11} h^{-1}M_{\sun}$ for
high-density early type galaxies at $L/L_*=1.1$ indicates a virial
radius of $r = 288$ \hkpc, which combined with the
\cite{2002ApJ...580L..11N} result suggests no dark matter beyond 10 per
cent of the virial radius for an equivalent halo in the field, or 85
per cent of the halo stripped (rather than our assumption of no matter
beyond 40 percent of $r_{vir}$, or half the mass stripped).  Tidal
stripping beyond $28$ \hkpc{} would give a much more suppressed signal
on scales $\sim 28$ to $200$ \hkpc{} (above which the group/cluster
contribution dominates) than what we see here.  However,
we note that those results are for a small number of massive clusters
(five), whereas 
our results are an average over all group- and cluster-sized halos.
Furthermore, our results average over all 
separations from the cluster center, whereas the results from
\cite{2002ApJ...580L..11N} are all within 50-100 arcsec, or a
characteristic scale of 300 \hkpc{}, significantly smaller than the
virial radius of a massive cluster, and therefore in the regime where
the tidal stripping is expected to be more efficient, so it is not 
possible to draw quantitative conclusions from this comparison.
\cite{2004A&A...422..407G} also found, using a weak lensing analysis
of a single supercluster, that for early types, the
typical scale of the dark matter halos of satellites decreased from
about 300 kpc on the outskirts of clusters to 200 kpc near the cluster
cores; this result is consistent with our assumed model of
tidal stripping, and with our findings.

A number of previous studies have attempted to determine the relationship
between mass and light.  \cite{2002MNRAS.335..311G} found a roughly
power-law $M(L) = M_* (L/L_*)^{\beta}$ in each band using a halo model
analysis of g-g weak lensing data in 
each band from the SDSS, but with a much smaller galaxy sample, and
different methods of calibrating the shear and the redshift
distributions than are used here.  Their 
results in the $r$-band were $M=(8.96\pm 1.59)\times
10^{11}(L/L_*)^{1.51\pm 0.04}
h^{-1}M_{\sun}$, as 
determined using luminosity samples spanning the range $0.6<L/L_*<7$.
However, these results assumed $\alpha=0$; using a more reasonable
value of $\alpha \sim 0.2$ gives values of $M_*$ that are 10--20 per
cent lower.  Furthermore, that work defined the mass using
$200\rho_{crit}$ rather than $180\overline{\rho}$, which means that in
order to compare their $M_*$ with ours, we must increase their value of
$M_*$ by approximately 30 per cent.  Thus, from the combination of
these two effects, we increase their $M_*$ by 10 per cent, yielding $M_*
= (9.9\pm 1.7) \times 10^{11} h^{-1}M_{\sun}$ ($1\sigma$ error).  In
Table~\ref{T:massvslum}, we can see that for the 
luminosity bin closest to $L_*$, our central halo masses are
$(14.9_{-4.6}^{+5.0})$ and $(7.1_{-3.0}^{+2.8})\times 10^{11}
h^{-1}M_{\sun}$ for early- and late-types respectively.  So, using
the 35 per cent spiral fraction for this bin from
Table~\ref{T:lumsplit}, we get an average halo mass of $(12.2_{-2.5}^{+2.4})\times
10^{11} h^{-1}M_{\sun}$ ($2\sigma$), which is 20 per cent higher than
but statistically consistent with the value
from \cite{2002MNRAS.335..311G}.  We also compare against their $M_*$ values
for early- and late-types in the $r$-band, also corrected upwards by
10 per cent, which are $(11.8\pm 2.8)$ and $(3.6\pm 2.3)\times 10^{11}
h^{-1}M_{\sun}$ (where the fits allowed $\alpha$ to vary, unlike the
above estimates, and fixed $\beta$ to 1.51), which is also lower than
but consistent
with our results. In Fig.~\ref{F:lumtrends}, the top panel includes as
dashed lines the $M(L)$ for early and late types from
\cite{2002MNRAS.335..311G}; as shown, the fit, while not perfect, is
not totally ruled out by our $2\sigma$ errors.  In particular, for
late types, the slope appears reasonable but the amplitude slightly
low compared to our results, whereas for early types, the deviation
between our results and the fits is most prominent for high
luminosities, where the fits underestimate $M_{cent}$ (i.e., our
results are consistent with a steeper slope $\beta$).  We note that
there are a number of other differences in the modeling between this
paper and that one (e.g., the determination of $\alpha$ and a lack of
corrections for width of the central halo mass distribution in the
older work), and differences in calibration of the lensing
signal, so a more detailed comparison is not necessarily
useful or meaningful.

Next, we compare against the recent work by
\cite{2005ApJ...635...73H}.  A number of factors make direct
comparison difficult: the different average lens redshift (with
complications due to luminosity evolution),
the use of different passbands ($V$, $B$, and $R$ rather than SDSS
bands), the different definitions of virial mass, and the selection of
only isolated galaxies to avoid the need to fit for a non-central
term.  However, a 
number of their findings may be compared directly against ours.
First, they find $M \propto L^{1.5}$ in all three bands, similar to
\cite{2002MNRAS.335..311G}; the previous paragraph includes a
discussion of this result relative to ours as shown in
Fig.~\ref{F:lumtrends}.  Next, when accounting for luminosity evolution,
difference in passbands, and different definitions of mass,
\cite{2005ApJ...635...73H} find
that their masses are about 25 per cent higher than those from
\cite{2002MNRAS.335..311G} when comparing against the $g$-band results, but
still statistically consistent. This result implies that our results
for the masses are consistent with those from
\cite{2005ApJ...635...73H}, since our masses are also slightly higher
than those from \cite{2002MNRAS.335..311G}.  Finally,
\cite{2005ApJ...635...73H} find conversion efficiencies a factor
of two higher for late types than for early types, 33 per cent versus
14 per cent.  This difference in morphology appears to be consistent
with our results 
(see Fig.~\ref{F:smtrends}) for stellar masses larger than
$10^{11}M_{\sun}$ or $L \ge L_*$; for lower stellar masses or
luminosities, we lack
statistical precision to make concrete statements.

The conditional luminosity function fits to 2dF data from \cite{2003MNRAS.340..771V} yield $M(L)$ that is
quite similar to 
that given in \cite{2004MNRAS.353..189V} based on empirical models,
which is $M \propto L^{0.25}$ at low mass (or 
luminosity), and $M \propto L^{3.6}$ at the high mass end, where the
luminosity is that of the BCG alone (i.e. not all the cluster galaxies
combined) and the mass is that of the full cluster.  While we are
unable to constrain the power-law slope very well at low luminosities
because of the errors, the three brightest luminosity bins with (L5f,
L5b, L6f) give $M \propto L^{2.7\pm 0.6}$ (95 per cent CL), with
exponent about
$3\sigma$ below the \cite{2004MNRAS.353..189V} model predictions.  
\cite{2004bdmh.confE..41V} figure 1 also shows that the 
conversion efficiency is highest for $L_{*}$ galaxies, with
rapidly increasing $M/L$ for 
lower and higher masses, consistent with our results. 

%
Analytical models of the Milky Way
\citep{2002ApJ...573..597K} predict halo 
masses of $7\times 10^{11} h^{-1}M_{\sun}$, where the mass is defined
as that within the radius within which the average density is equal to
$340\overline{\rho}$.  To compare against our results, their mass
estimate must be
increased by 15 per cent, giving $8\times 10^{11}h^{-1}M_{\sun}$,
which we compare 
against our result for $L_*$ late-type galaxies, $M_{cent} =
(7.1_{-3.0}^{+2.8})\times 10^{11} h^{-1}M_{\sun}$ (95 per cent CL).
Alternatively, we note that their table 2 suggests a total stellar
mass of $6\times 10^{10}M_{\sun}$, which is typical of $L_*$ galaxies
and which (according to our Table~\ref{T:massvssm}) gives a halo mass of
$M_{cent} = 14_{-7}^{+8} \times 10^{11} h^{-1}M_{\sun}$.  
Thus, it appears that this analytical model of the Milky Way is
consistent with our results at the 95 per cent CL.

It is also worthwhile to compare against the lensing signal from
$N$-body simulations.  In this case, we compare against the fit results
from \cite{2005MNRAS.362.1451M}, which showed the lensing signal for
three luminosity bins from simulations described in
\cite{2004ApJ...614..533T}, both with and without scatter in the 
mass-luminosity relationship: L3, L4, and L5 (a bin one
magnitude wide that includes L5f and L5b).  Without scatter, the
values of $M_{cent}$ (mean) in each bin were $5.1$, $18$, and
$132\times 10^{11} h^{-1}M_{\sun}$; with scatter, they were $7.5$,
$29$, and $117\times 10^{11} h^{-1}M_{\sun}$.  For this paper, if we
combine the results in each bin (averaging the results with different
morphologies using the spiral fraction for weighting purposes), we get
$4.3 \pm 2.5$, $12\pm 3$, and $(82\pm 15)\times
10^{11}h^{-1}M_{\sun}$.  Therefore, the results from simulations seem
to give somewhat higher masses than are found in the real data, though
just within the 95 per cent confidence intervals, except for L5.

Finally, we compare against the baryonic mass function determination
by \cite{2005astro.ph..2517R} that uses a variety of data sources.  In
that work, the fraction of baryons in galaxies is estimated to be
$\sim 10$ per cent, and of those, $\sim 80$ per cent are in stars (with the
star formation efficiency varying by morphological type). Hence, the
\cite{2005astro.ph..2517R} results suggest average $\eta$ values of 0.08.
While our peak values of $\eta$ (around $10^{11} \hMsun$) are higher
than this, the abundance of galaxies at much lower stellar mass (for
which our central values of $\eta$ are lower but poorly constrained)
suggests that our results and those of \cite{2005astro.ph..2517R} are
consistent within the errors.  Furthermore, their results and ours
both suggest the trend of decreasing $\eta$ with increasing stellar
mass above $10^{11} \hMsun$.  They attribute this trend to feedback
from AGNs which would tend to lower the baryonic fraction in high
mass galaxies (e.g., \citealt{1998A&A...331L...1S}).

\subsection{Satellite fractions}

\begin{figure}
\includegraphics[width=3.0in]{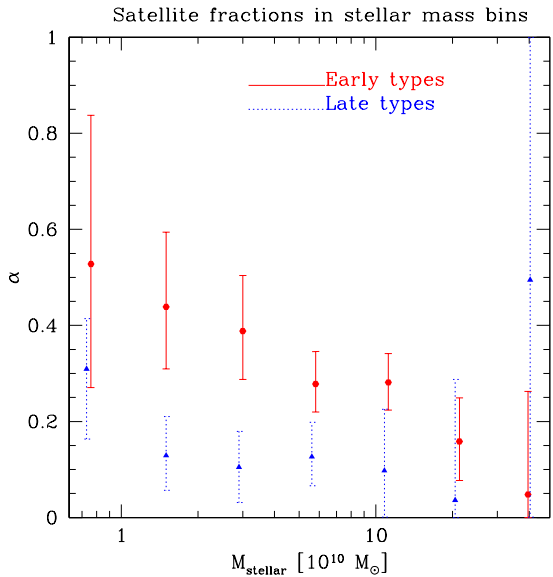}
\caption{\label{F:smalpha}Satellite fractions as a function of
  stellar masses for samples split into early (red hexagons, solid
  lines) vs. late (blue triangles, dotted lines) types.
  Errorbars shown are 
  95 per cent confidence intervals.}  
\end{figure}

\begin{figure}
\includegraphics[width=3.0in]{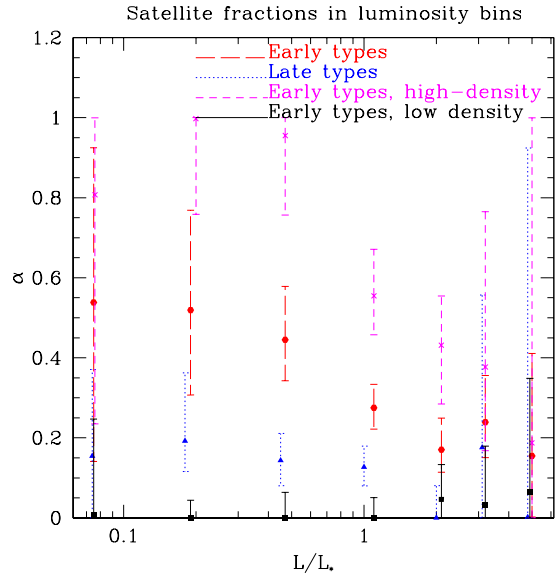}
\caption{\label{F:lumalpha}Satellite fractions as a function of
  luminosity for samples split into early (red hexagons, long-dashed
  lines) vs. late (blue triangles, dotted lines) types, and early
  types split into high-density (magenta crosses, dashed lines) and
  low-density  (black squares, 
  solid lines) samples.  Errorbars shown are
  95 per cent confidence intervals.}  
\end{figure}

Here we present results for the satellite fractions as a function of
luminosity,  stellar mass, and morphology.
Table~\ref{T:smalpha} and Fig.~\ref{F:smalpha} show the best-fit
$\alpha$ in stellar mass bins 
with 95 per cent confidence intervals from the bootstrap, for early and late types
separately.  Table~\ref{T:lumalpha} shows the best-fit value of $\alpha$ in
luminosity bins with 95 per cent confidence intervals, for early and
late types, and for early-types split by local density, and
Figure~\ref{F:lumalpha} shows these results as well.  We note that
$M_{cent}$ and $\alpha$ have a cross-correlation coefficient of about
$-0.7$ from the fits, because the choice of $M_{cent}$ determines the
mass at which \avgnm{} becomes proportional to $M$, which has a
significant effect on the satellite contribution.

\begin{table}
\caption{\label{T:smalpha}The satellite fraction determined for
  stellar mass subsamples separately for early- and late-type galaxies.
  All confidence intervals are 95 per cent.}
\begin{tabular}{lcc}
\hline\hline
$\langle M_{stellar} \rangle$ & $\alpha$ & $\alpha$ \\
 $[10^{10} M_{\sun}]$ & Early-types & Late-types \\
\hline
0.76 & $0.53_{-0.26}^{+0.31}$ & $0.31_{-0.15}^{+0.10}$ \\
1.5 & $0.44_{-0.13}^{+0.16}$ & $0.13_{-0.07}^{+0.08}$ \\
3.0 & $0.39_{-0.10}^{+0.12}$ & $0.10_{-0.07}^{+0.07}$ \\
5.7 & $0.28_{-0.06}^{+0.07}$ & $0.13_{-0.06}^{+0.07}$ \\
11.1 & $0.28_{-0.06}^{+0.06}$ & $0.10_{-0.10}^{+0.13}$ \\
21.0 & $0.16_{-0.08}^{+0.09}$ & $0.04_{-0.04}^{+0.25}$ \\
40.0 & $0.05_{-0.05}^{+0.21}$ & $0.47_{-0.47}^{+0.53}$ \\
\hline\hline
\end{tabular}
\end{table}
\begin{table*}
\caption{\label{T:lumalpha}The satellite fraction determined for
  luminosity subsamples separately for early- and late-type galaxies,
  and for early-type galaxies split into low-density and high-density  samples.
  All confidence intervals are 95 per cent.}
\begin{tabular}{lcccc}
\hline\hline
$\langle L/L_*\rangle$ & $\alpha$ & $\alpha$ & $\alpha$ & $\alpha$ \\
 & Early & Late & Early & Early \\
 & (all) & (all) & (low density) & (high density) \\
\hline
0.075 & $0.54_{-0.40}^{+0.39}$ & $0.15_{-0.15}^{+0.22}$ & $0.006_{-0.006}^{+0.24}$ & $0.81_{-0.57}^{+0.19}$ \\
0.19 & $0.52_{-0.21}^{+0.25}$ & $0.19_{-0.08}^{+0.17}$ & $(1.2_{-1.1}^{+442})\times 10^{-4}$ & $1.00_{-0.24}^{+0.00}$ \\
0.47 & $0.44_{-0.10}^{+0.13}$ & $0.14_{-0.06}^{+0.07}$ & $(3.1_{-1.9}^{+6376})\times 10^{-5}$ & $0.96_{-0.20}^{+0.04}$ \\
1.1 & $0.27_{-0.05}^{+0.06}$ & $0.13_{-0.05}^{+0.05}$ & $(9.3_{-8.5}^{+5079})\times 10^{-5}$ & $0.55_{-0.10}^{+0.12}$ \\
2.1 & $0.17_{-0.06}^{+0.08}$ & $(2.2_{-2.1}^{+804})\times 10^{-4}$ & $0.046_{-0.046}^{+0.087}$ & $0.43_{-0.15}^{+0.12}$ \\
3.2 & $0.24_{-0.09}^{+0.12}$ & $0.18_{-0.18}^{+0.38}$ & $0.03_{-0.03}^{+0.15}$ & $0.38_{-0.21}^{+0.39}$ \\
4.9 & $0.15_{-0.15}^{+0.26}$ & $0.0017_{-0.0016}^{+0.92}$ & $0.07_{-0.07}^{+0.28}$ & $0.19_{-0.19}^{+0.81}$ \\
\hline\hline
\end{tabular}
\end{table*}

We discuss both Figs.~\ref{F:smalpha} and~\ref{F:lumalpha} together.
For both the splits by luminosity and by stellar mass, the trend is
for $\alpha$ for early types to decrease slightly with $L$ or
$M_{stellar}$, with central values around 0.5 at $M_{stellar} \sim
0.8\times 10^{10} M_{\sun}$ or $L/L_*\sim 0.1$, decreasing to a
central value of 0.1 (upper limit $\sim 0.2$) for $M_{stellar} \ge 20
\times 10^{10} M_{\sun}$, or 0.2 (upper limit $\sim 0.35$) for
$L/L_* >2$.  We remind the reader that these numbers do not limit the
number of early-type lenses that are in groups and clusters, because a
galaxy that is the BCG of a group/cluster (of which we expect many in
L5 and L6) will {\it not} be included as a satellite.  Hence, it is
possible that the number of 
early-type galaxies in groups or clusters is actually constant with
$L$ or $M_{stellar}$ when the number that are BCGs is accounted for;
unfortunately, we cannot test this hypothesis using our halo model formalism.

For early types, we also have done a simultaneous split by luminosity and
density.  As shown, for $L/L_* < \sim 2$, this split seems to do an
excellent job of isolating those early type galaxies that are in groups
and clusters.  For example, for L3 ($L/L_* \sim 0.5$), we see that the
central value of $\alpha$ for the full sample of early type galaxies
is 0.44, for the low-density elliptical sample is consistent with
zero, and for the high-density elliptical sample is 0.96.  Since the
low- and high-density samples are determined by splitting at the
median environment estimate, these results for satellite fractions are
fully consistent with each other.  For all luminosity bins, similar
consistency relationships are satisfied within the noise.  We expect
that for higher luminosities, many of the high-density sample are BCGs
of groups and clusters, which may account for the discrepancy noted in
central halo masses in the previous section, and for the decline of
satellite fraction with luminosity for this sample.

For late types, there is no suggestion of a statistically significant
change in the group/cluster fraction with luminosity or stellar mass,
with central values $\sim 0.10$--$0.15$ and upper bounds typically 0.25--0.3.
The lower satellite fraction for late types than for early types is
consistent with works cited previously.

The overall trend of satellite fraction decreasing with mass is also
observed with simulations and semi-analytic galaxy formation models of
\cite{2004astro.ph..8564Z}.  

We compare these results against derived satellite fractions from
\cite{2005ApJ...630....1Z} for L2, L3, and L4.  They find that 10--30
per cent of blue galaxies are satellites  independent of luminosity,
consistent with our results.  Of the red galaxies, they find satellite
fractions of 0.72, 0.54, 0.35 for L2, L3, and L4 respectively, so the
trend of $\alpha$ decreasing with luminosity is found in both their
autocorrelation analysis and our lensing analysis.  However, our
values of $\alpha$ for these bins are $0.52_{-0.21}^{+0.25}$,
$0.44_{-0.10}^{+0.13}$, and $0.27_{-0.05}^{+0.06}$, about 30 per cent
lower than their results, though for L2 and L3 their values of
$\alpha$ lie within our 95 per cent confidence intervals; for L4, they
lie slightly outside our 95 per cent intervals, but when the errors on
their estimates are included, it is not clear that the discrepancy is
significant.  They also find that the average host halo mass for L2
galaxies (i.e., the mass of the full cluster if they are in a cluster,
or the halo mass of field galaxies, which are in the minority) is $2.5
\times 10^{14} h^{-1}M_{\sun}$.  We have found
\citep{2005MNRAS.362.1451M} that the lensing signal on group and
cluster scales is actually dominated by less massive halos,
in the mass range $10^{13}$--$10^{14}$; if the average host halo mass
was as high as the value given in \cite{2005ApJ...630....1Z}, then
considering the large satellite fraction in this bin, the lensing
signal would be too large to be consistent with observations on 500
\hkpc{}-- 1 \hmpc{} scales (increasing by more than a factor of two
from its current value).

\subsection{Robustness of best-fit parameters}

In order to determine the robustness of these results, we consider
which modeling assumptions may affect them.  The first
effect we consider is that of $\avgnm$ which has been modeled
here as a power-law $\avgnm \propto M$ for $M>3 M_{cent}$
and $\propto M^2$ for $M<3M_{cent}$, so $3M_{cent}$ marks the point
below which it falls off rapidly.  This choice was justified in
\S\ref{S:hm}, but we must consider the effects of this choice on our
results for the best-fit satellite fractions.  In particular, in
\cite{2005MNRAS.362.1451M}, we found that the power-law exponent
$\epsilon$ of \avgnm{} for $M>3M_{cent}$ was almost completely
degenerate with $\alpha$.  The two parameters essentially arranged
themselves to preserve the amount of signal coming from halos in the
mass range $10^{13}$--$10^{14}$ $h^{-1}M_{\sun}$ (higher mass halos
are not important because they are killed off by the exponential in
the mass function $dn/dM$, and lower mass halos do not contribute a
significant amount of signal).  Thus, our satellite fractions have the
potential to differ significantly from the real one if $\epsilon$ is incorrect.
  
\cite{2003MNRAS.340..771V} find using data from 2 Degree Field Galaxy
Redshift Survey \citep[2dFGRS, ]{2001MNRAS.328.1039C} that
$\avgnm$ has a 
different power-law dependence for spirals (shallower) and ellipticals
(steeper), with the
net result being that the bright luminosity bins, which are dominated
by ellipticals, show a single power-law, but fainter luminosity bins
are best described by a combination of the two power laws, with the
spiral one dominating at low mass and the elliptical one dominating at
high mass.  This finding is reasonable in light of the fact that
spirals are known to dominate in the field and ellipticals in
clusters.  However, it does mean that our assumption of $\epsilon=1$
for both early- and late-types may complicate our analysis, since it
may have caused an overestimate of $\alpha$ for early-types and
underestimate for late-types, which is exactly in the direction of the
difference we measured (i.e., lower satellite fractions
for late types).  

As noted in \cite{2003MNRAS.340..771V}, which seems to suggest
$\epsilon \sim 0.8$ for ellipticals 
and $\sim 0.6$ for spirals using the 2dFGRS $b_J$ band data (in
agreement with another $b_J$ band study using APM data,
\citealt{2001ApJ...546...20S}, which found $\epsilon \sim 0.8$
overall), we do not expect $\epsilon$ to be the same when determined
using data selected by absolute luminosity in different bands, so in
order to estimate its value for early- and late-types and its
possible luminosity evolution, we turn to \cite{2005ApJ...630....1Z},
which uses SDSS samples selected in the $r$-band and a halo model
analysis of $\xi_{gg}(r_p)$ in order to determine this
parameter.\footnote{Note that both \cite{2005ApJ...630....1Z} and
  \cite{2004astro.ph..8564Z} call this parameter $\alpha$ (i.e.,
  the symbol we use for the satellite fraction), rather than
  $\epsilon$.}  The right-hand side of figure 18 in that paper shows
the best-fit value of $\epsilon$ for samples selected by luminosity
{\it thresholds} rather than 1-magnitude wide bins; as shown, the
value of $\epsilon$ is $\sim 0.9$ for $M_r < -18$ samples, rising
slowly to $\epsilon \sim 1.3$ for $M_r < -20.5$ samples, then rising
sharply to $\epsilon \sim 2$ above that.  This trend may reflect the
relatively higher fraction of early types in the brighter
luminosity-threshold samples.  The lower right panel of their figure
22 shows the best-fit $\avgnm$ for early- and late-types, from which
we deduce values of $\epsilon \sim 1.65$ and $\sim 1.10$,
respectively.  We note that \cite{2004astro.ph..8564Z}, on the other
hand, find $\epsilon \sim 1$ for all samples with the same data,
reflecting a difference in the modeling.  The difference may be
attributed to the fact that \cite{2005ApJ...630....1Z} lack a lower
mass cut-off for their \avgnm, whereas this work and
\cite{2004astro.ph..8564Z} do include a lower mass cutoff.  Because of
the lack of cut-off in \cite{2005ApJ...630....1Z}, their value of
$\epsilon$ must necessarily be quite high for the brightest samples in
order to limit the contribution from lower mass halos.  We note that
\cite{2005astro.ph.12234C} have explicitly
tested this hypothesis by fitting for $\epsilon$ with and without a
lower mass cutoff, and 
found that the lack of lower mass cutoff can increase the best-fit
$\epsilon$ by as much as 50 per cent.

We consider the effect of changing $\epsilon$ on the best-fit values of
$\alpha$ for
early and late types separately by calculating an analytic correction assuming
the \cite{2005ApJ...630....1Z} values of $\epsilon$ for those samples
independent of luminosity 
or stellar mass; this
correction requires that we compute the non-central lensing signal for
satellites residing in $10^{13}$--$10^{14}$ 
$h^{-1}M_{\sun}$ halos as the product of $\alpha$ times an integral
involving \avgnm{}, and requiring that it be preserved when we change
$\epsilon$, thus telling us the new value of $\alpha$ as well.  We
note that the correction factor is a function of central 
halo mass because, for high values of central halo mass, the cutoff
$3M_{cent}$ will be within our mass range of interest, so changing
$\epsilon$ will have less of an effect.  For both 
stellar mass and luminosity bins, the results of this calculation indicate that
for late types, changing $\epsilon$ from our assumed value of 1.0 to
the \cite{2005ApJ...630....1Z} value of 1.1 requires that we decrease
$\alpha$ by 4 per cent of its best-fit value, which is significantly
less than the statistical error on this value for any stellar mass
bin.  For stellar mass samples, for early types, changing $\epsilon$
from 1.0 to 1.65 requires that we decrease $\alpha$ by 21 per cent of
its original value for
the 5 lowest stellar mass bins, 12 per cent for sm6, and no decrease
for the largest stellar mass bin (due to its high central halo mass).
For luminosity samples, for early types, it is to decrease
$\alpha$ by 21 per cent for the 5 lowest bins ($-17 \ge M_r > -21.5$),
by 12 per cent for $-21.5 \ge M_r > -22$, and no decrease for $-22 \ge
M_r > -22.5$.  Thus, the value of $\alpha$ for late types is nearly
unchanged, and 
for early types, the limiting value at lower stellar mass becomes
$0.4$ rather than $0.5$.  We note that for the split by environment,
since we only use early types, the same correction factors apply as
for the full early type sample, so while the actual values of
$\alpha$ must be lowered, this correction does not affect the
consistency relationship between the satellite fractions in the low-
and high-density early-type samples relative to the full early-type
samples.  Finally, we remind the reader that since the
\cite{2005ApJ...630....1Z} values of $\epsilon$ are higher than our
assumed value in part due to a significant difference in modeling, and
such high values of $\epsilon$ are not consistent with N-body simulation
results when using a model similar to ours,
these corrections are actually quite conservative.

Next, we consider the fact that the expected signal depends on the
radial distribution of satellites within groups and clusters.  As
mentioned in \S\ref{S:hm}, this distribution is not well known, and is
assumed here to be an NFW profile with the same concentration
parameter as the DM.  Variations in $c_g$ cause the
noncentral signal to peak at different characteristic radii as shown
in~\cite{2002MNRAS.335..311G}, thus affecting primarily the shape of
the signal rather than its amplitude.  Unfortunately, since in most
cases the values of $\alpha$ are $\sim 0.2$--$0.3$, we are not highly
sensitive to $c_g$ and cannot place much of a constraint on it. 

Several observational results have suggested that within clusters, red
galaxies are more centrally concentrated than blue ones
\citep{1974ApJ...194....1O,1977ApJ...215..401M,1980ApJ...236..351D,1998A&A...331..439A}.
Due to a relatively low sensitivity to $c_g$, we do not fit for it,
but see what happens to the fit $\chi^2$ and to the best-fit
$M_{cent}$ and $\alpha$ if we use $c_g = 2c_{dm} \sim 24$ for early-types, and
$c_g = 0.5c_{dm} \sim 6$ for late-types.  We do this comparison only for L3,
L4, L5f, and L5b, since these are the samples with the greatest
statistical power.  

For early types, we find that because increasing $c_g$ moves the
non-central contribution of the signal to 
relatively small scales, this change actually affects the central halo
mass $M_{cent}$ more than it affects the satellite fraction, with the
tendency being to decrease the halo mass slightly to compensate for
the higher non-central signal on $\sim 100-200$ \hkpc{} scales.  This
change only affects the best-fit $\chi^2$ by $\sim 1$, with the change
being in different directions for the different luminosity bins.  The
masses in L3, L4, L5f, and L5b decrease to 60, 59, 87, and 86 per cent
of their values from the fits with $c_g=c_{dm}$, and the satellite
fraction changes by -0.01, -0.02, -0.01, and -0.01 (absolute value,
not per cent of original).  Hence, the changes to $\alpha$ for early
types are well
within the statistical errorbars, and we conclude that best-fit values
are relatively robust to uncertainties in $c_g$.  The changes to the
best-fit values of $M_{cent}$ are at most $1\sigma$ (for L3) and
usually somewhat less than that, and thus we conclude that this
parameter is also not sensitive to changes in $c_g$ within our
statistical errorbars, though with more data this statement may no
longer be true.

For late types, the decrease from $c_g=c_{dm}$ to $c_g=0.5c_{dm}$
decreased the best-fit $\chi^2$ by at most 1, and had less of an
effect on the best-fit $M_{cent}$ because it shifted the non-central
contribution to higher radius.  The masses in L3, L4, L5f, and L5b
increased by 9, 5, 1, and 15 per cent of their original values, well
within the $1\sigma$ errors, and $\alpha$ increased by 0.02, 0.02,
0.00, and 0.00, also well within the errors.  We thus conclude that
uncertainty in the distribution of satellites within groups and
clusters is not a significant source of systematic uncertainty in our
estimates of satellite fractions and central halo masses. 

Another source of uncertainty in these estimates is our modeling of
the h-h term, which affects the signal on large scales, and thus can
change the best-fit $\alpha$ and (through their degeneracy)
$M_{cent}$.  Fortunately, since the h-h term is small on $r < 2$
\hmpc{} scales, we find that neglecting it entirely changes the
best-fit parameters by less than $1\sigma$.

Finally, we remind the reader that due to the assumption of a
universal IMF when deriving stellar mass estimates, these estimates
may need to be rescaled by a constant factor of up to 30 per cent
(conservatively).
The main results that are affected by this rescaling are the limits
that we have placed on conversion efficiencies.  All other mass-related
results, such as trends in halo masses with stellar mass, and
comparisons between different morphology or density samples at
constant stellar mass, are unaffected.

\section{Conclusions}\label{S:conclusions}

In this work, we have used halo model analysis of the galaxy-galaxy
weak lensing signal in order to observe trends in the relationship
between stellar masses and halo masses, and luminosity and halo mass,
treating samples based on morphology separately.  We have also studied
ellipticals in low- and high-density regions separately.

As a result, we have come to a number of conclusions
related to average halo masses.   First, the $M_{stellar}/M_{cent}$
ratio is highest for $M_{stellar}\sim 10^{11}M_{\sun}$, with a peak
conversion efficiency of roughly 14 per cent for ellipticals ($>11$
per cent at 95 per cent CL, statistical error only), 
although we cannot exclude even larger values at low luminosity or
stellar mass where 
the lensing signal is weak.  
The corresponding number for spirals reaches 35 per cent at the maximum, 
but with a larger measurement error, which allows it to be as low as 
18 per cent (95 per cent CL).  These results imply a factor of two or more
difference in conversion efficiency between typical spirals and
ellipticals above stellar mass of $10^{11}M_{\sun}$, whereas below
this stellar mass, we find no inconsistency between the conversion
efficiencies of early and late types, implying that stellar mass is a
good tracer of halo mass in this regime.
Note that since we are only including central galaxy stallar mass in
the analysis,  
our conversion efficiencies should be viewed as lower limits, and also
serve as lower limits on star formation efficiency. 
In practice, for a $10^{12}M_{\sun}$ halo, the 
central galaxy likely contains most of the stars and the contribution 
from satellites is negligible, while for 
a $10^{15}M_{\sun}$ halo other galaxies outside the halo center contribute 
significantly to the total stellar content of a cluster. 
Thus while the upper limits to the conversion efficiency 
are uncertain both because of large observational errors
and because of modeling 
uncertainties, a lower limit of about 10 per cent 
(95 per cent CL) is robust for both spirals and ellipticals.
Similarly, $M/L$
reaches a minimum for $L\sim L_*$ galaxies of $41_{-17}^{+16}
M_{\odot}/L_{\odot}$ for late types, or for $L \sim L_*/2$ of
$51_{-37}^{+51}$ for early types (95 per cent CL). Below these values
of stellar mass or luminosity, both of those quantities trace halo
mass well, meaning that halo masses for early and late types were
consistent.  At higher stellar mass or luminosity, the early types
have a larger central halo mass, likely reflecting the fact that they
tend to reside at the center of clusters, unlike late types.

We also have a number of conclusions regarding satellite fractions.
The satellite fraction for late types tends to be 10-15 per cent
regardless of luminosity.  For early types, the situation is more
complex, with a satellite fraction of $\sim 40-50$ per cent at low
luminosity, decreasing to about 20 per cent at high luminosity.  
These trends are consistent with those based on auto-correlation 
analysis \citep{2005ApJ...630....1Z}, though the actual values of
satellite fraction for early types are slightly different, possibly
reflecting differences in modeling. 
Our fit results for $L<\sim L_*$ early types indicate that our density
estimate is highly efficient at isolating a nearly pure satellite
sample in this regime, which can then be used in a future work that
will study tidal 
stripping and the radial distribution of satellites within groups and
clusters. Current results suggest that tidal stripping is not 
completely efficient in removing the dark matter from the satellites, since 
the satellite sample shows plenty of lensing signal at small (50--100
\hkpc) scales.  

\section*{Acknowledgements}

US is supported by a fellowship from the
David and Lucile Packard Foundation,
NASA grant NAG5-11489 and NSF grant CAREER-0132953.  CH is supported
in part by NSF PHY-0503584 and by a grant-in-aid from 
the W. M. Keck Foundation. We thank Nikhil
Padmanabhan for useful discussions related to this work, and the
anonymous referee for useful comments.

Funding for the creation and distribution of the SDSS Archive has been
provided by the Alfred P. Sloan Foundation, the Participating
Institutions, the National Aeronautics and Space Administration, the
National Science Foundation, the U.S. Department of Energy, the
Japanese Monbukagakusho, and the Max Planck Society. The SDSS Web site
is http://www.sdss.org/. 

The SDSS is managed by the Astrophysical Research Consortium (ARC) for
the Participating Institutions. The Participating Institutions are The
University of Chicago, Fermilab, the Institute for Advanced Study, the
Japan Participation Group, The Johns Hopkins University, the Korean
Scientist Group, Los Alamos National Laboratory, the
Max-Planck-Institute for Astronomy (MPIA), the Max-Planck-Institute
for Astrophysics (MPA), New Mexico State University, University of
Pittsburgh, University of Portsmouth, Princeton University, the United
States Naval Observatory, and the University of Washington.

\bibliography{../BibTeX/apjmnemonic,../BibTeX/cosmo,../BibTeX/cosmo_preprints}
\bibliographystyle{mn2e}

\end{document}